\documentclass[journal]{IEEEtran}
\usepackage[T1]{fontenc}
\usepackage[latin9]{inputenc}
\usepackage{float}
\usepackage{amsmath}
\usepackage{amssymb}
\usepackage{stmaryrd}
\usepackage{stackrel}
\usepackage{graphicx}
\usepackage[unicode=true,
 bookmarks=true,bookmarksnumbered=true,bookmarksopen=true,bookmarksopenlevel=1,
 breaklinks=false,pdfborder={0 0 0},pdfborderstyle={},backref=false,colorlinks=false]
 {hyperref}
\hypersetup{pdftitle={Your Title},
 pdfauthor={Your Name},
 pdfpagelayout=OneColumn, pdfnewwindow=true, pdfstartview=XYZ, plainpages=false}

\makeatletter

\floatstyle{ruled}
\newfloat{algorithm}{tbp}{loa}
\providecommand{\algorithmname}{Algorithm}
\floatname{algorithm}{\protect\algorithmname}

\usepackage[caption=false,font=footnotesize]{subfig}

\makeatother

\begin{document}
\title{Exploiting Spatial Multiplexing Based on Pixel Antennas: An Antenna
Coding Approach}
\author{Zixiang Han,~\IEEEmembership{Member,~IEEE,} Shanpu Shen,~\IEEEmembership{Senior Member,~IEEE,}
and~Ross Murch,~\IEEEmembership{Fellow,~IEEE}\thanks{Manuscript received; This work was funded by the Science and Technology
Development Fund, Macau SAR (File/Project no. 001/2024/SKL). \textit{(Corresponding
author: Shanpu Shen).}}\thanks{Z. Han is with Future Research Lab, China Mobile Research Institute,
Beijing, 100053, China. (e-mail: hanzixiang@chinamobile.com)}\thanks{S. Shen is with the State Key Laboratory of Internet of Things for
Smart City and Department of Electrical and Computer Engineering,
University of Macau, Macau, China. (e-mail: shanpushen@um.edu.mo)}\thanks{R. Murch is with the Department of Electronic and Computer Engineering
and the Institute for Advanced Study, The Hong Kong University of
Science and Technology, Clear Water Bay, Kowloon, Hong Kong. (e-mail:
eermurch@ust.hk)}}
\maketitle
\begin{abstract}
An antenna coding approach for exploiting the spatial multiplexing
capability of pixel antennas is proposed. This approach can leverage
additional degrees of freedom in the beamspace domain to transmit
more information streams. Pixel antennas are a general reconfigurable
antenna design where a radiating structure with arbitrary shape and
size can be discretized into sub-wavelength elements called pixels
which are connected by radio frequency switches. By controlling the
switch states, the pixel antenna topology can be flexibly adjusted
so that the resulting radiation pattern can be reconfigured for beamspace
spatial multiplexing. In this work, we introduce the antenna coder
and pattern coder for pixel antennas, provide a multiple-input multiple-output
(MIMO) communication system model with antenna coding in the beamspace
domain, and derive the spectral efficiency. Utilizing the antenna
coder, the radiation pattern of the pixel antenna is analyzed and
efficient optimization algorithms are provided for antenna coding
design. Numerical simulation results show that the proposed technique
using pixel antennas can enhance spectral efficiency of 4-by-4 MIMO
by up to 12 bits/s/Hz or equivalently reduce the required transmit
power by up to 90\% when compared to conventional MIMO, demonstrating
the effectiveness of the antenna coding technique in spectral efficiency
enhancement and its promise for future sixth generation (6G) wireless
communication.
\end{abstract}


\section{Introduction}

\IEEEPARstart{M}{ultiple-input} multiple-output (MIMO) antennas play
a critical role in modern mobile networks \cite{1353475}. Leveraging
the spatial degrees of freedom (DoF) brought by half-wavelength spaced
antenna arrays, the spectral efficiency of wireless communication
systems have been greatly enhanced \cite{1310320}. However, antennas
in conventional MIMO systems, such as the active antenna unit (AAU)
of base stations, have fixed configuration and characteristics, including
radiation patterns and polarizations, making the spectral efficiency
of conventional MIMO systems bounded by its array configurations \cite{TS38803}.
To satisfy the demands of higher data rate in future sixth generation
(6G) mobile network, novel antenna technology needs to be developed
and investigated to provide extra DoF and break through the performance
limits.

Pixel antennas are a promising reconfigurable antenna technology that
can provide extra DoF to design and enhance wireless communication
systems. The concept of pixel antennas is that the antenna radiating
structure with arbitrary shape and size can be discretized into sub-wavelength
elements denoted as pixels, and adjacent pixels can be connected or
disconnected by using radio frequency (RF) switches such as positive-intrinsic-negative
(PIN) diodes \cite{1367557}\nocite{Tang2021}\nocite{Zhang2022}\nocite{Tang2023}-\nocite{Zhang2022a}\nocite{9785489}\nocite{9852001}\nocite{10025410}\nocite{Zhang2021}\cite{Zhang2020}.
By controlling the switch states, the pixel antenna topology can be
flexibly adjusted so that the resulting antenna characteristics such
as radiation pattern can be reconfigured for enhanced wireless transmission.
Pixel antennas are a general reconfigurable antenna design approach
that is suitable for both the base station and user equipment and
a variety of pixel antennas have been designed. One type of pixel
antenna design uses a grid of pixels to implement beam-steering where
single or multiple beams can be excited and steered in the full three-dimension
(3D) space, while avoiding the high insertion loss and large footprint
of using phase shifters \cite{Tang2021}\nocite{Zhang2022}\nocite{Tang2023}-\nocite{Zhang2022a}\cite{9785489}.
Pixel antennas can also be used to design reconfigurable intelligent
surfaces to control the phase of the reflected wave \cite{9852001},
\cite{10025410}. In addition, the structure of pixelized surface
has been utilized to decouple compact MIMO antennas so that the mutual
coupling is reduced and ergodic channel capacity is maximized \cite{Zhang2021}.
It is also applied to control cross polarization ratio for polarization
improvement \cite{Zhang2020}. Nevertheless, these designs focus on
the antenna performance enhancement by pixel antennas at the level
of antenna hardware design, which overlooks the potential of pixel
antennas in providing additional DoF for enhanced wireless transmission
at the system level.

Another emerging technology is the fluid antenna system (FAS) \cite{Wong2021}.
Different from conventional antennas with fixed position, antennas
in FAS can move within a small range \cite{10653737}, so that FAS
can adapt to the channel environment by adjusting the antenna position.
An approach to realize FAS is utilizing the fluidity of liquid metal
or conductive fluid to achieve antenna movement \cite{9388928}, but
the tuning speed and accuracy of this approach are limited. Alternatively,
a more promising way to implement FAS is using reconfigurable pixel
antennas through carefully optimizing the switch states to mimic the
antenna position switching in a small linear space \cite{10740058}.
Leveraging the fluidity, the key performance of MIMO communication
systems, including multiplexing gain, diversity gain, beamforming
gain, spectral efficiency, as well as energy efficiency, can be enhanced
by FAS \cite{10130117}-\nocite{10303274}\nocite{10243545}\cite{10328751}.
The multiple access approach based on FAS, denoted as fluid antenna
multiple access (FAMA), has been proposed in \cite{10078147} to enhance
the multi-user communications. These studies on FAS have preliminarily
demonstrated the potential of pixel antennas in enhancing wireless
communication systems.

To fully exploit the potential of pixel antennas, a novel technique
denoted as antenna coding, which is based on pixel antennas, has recently
been proposed in \cite{shen2024antenna}. By controlling a binary
vector referred to as the antenna coder, which characterizes the switch
states, the radiation pattern of the pixel antenna can be optimized
through a beamspace channel representation to improve system-level
performance such as channel capacity. Thus, antenna coding further
generalizes the utilization of radiation pattern reconfigurability
in designing and enhancing wireless communication system and open
up new opportunities in exploring new DoF of pixel antennas in the
beamspace domain. Nevertheless, the previous work \cite{shen2024antenna}
primarily leverages the antenna coding technique to improve the channel
gain, through optimizing the radiation pattern to coherently add multiple
paths from different directions, while ignoring the possibility of
exploiting the spatial multiplexing to enhance the spectral efficiency
of the wireless communication system.

To overcome this limitation, in this work we exploit spatial multiplexing
by using an antenna coding approach based on pixel antennas to enhance
the spectral efficiency of wireless communication systems. Compared
to conventional MIMO systems with fixed antenna configurations, MIMO
systems implemented by pixel antennas with antenna coding enables
reconfigurable radiation patterns for each antenna, leveraging additional
DoF in the beamspace domain to transmit more information so as to
enhance spectral efficiency. The main contributions of this paper
are summarized as follows.

\textit{Firstly}, we exploit the spatial multiplexing capability of
pixel antennas by using an antenna coding approach to enhance the
spectral efficiency of MIMO systems. Using the reconfigurable pixel
antenna to take the place of the antennas with fixed configuration
in conventional MIMO system, the radiation patterns of each pixel
antenna can be flexibly adjusted according to antenna coders. This
allows transmitting additional information to be modulated by the
antenna coders and thus enhances the spectral efficiency of MIMO systems.

\textit{Secondly}, we derive the radiation pattern of pixel antennas
as a function of antenna coder. Based on multiport circuit theory,
we derive the relationship between the antenna coder and radiation
pattern of the pixel antenna. Accordingly, we propose the pattern
coder of pixel antenna which can be defined as linearly coding a set
of orthonormal basis radiation patterns to construct the radiation
pattern of the pixel antenna.

\textit{Thirdly}, we derive the MIMO system model with antenna coding.
Specifically, the beamspace channel representations is considered
to include the effect of antenna coding on the transmit radiation
patterns. The spectral efficiency of the MIMO system with antenna
coding is also analyzed. We show that the antenna coding technique
can increase spectral efficiency by transmitting additional information
associated with the radiation pattern selection of transmit antennas.

\textit{Fourthly}, we analyze the radiation pattern of the pixel antenna
by considering the mutual coupling strength between the antenna port
and pixel ports. The effective number of orthonormal basis radiation
patterns, i.e. effective aerial DoF (EADoF) in the beamspace domain,
for the pixel antenna is also analyzed. We also formulate the antenna
coding optimization problem to design a codebook to implement the
multiple orthonormal basis radiation patterns for spectral efficiency
enhancement. In addition, to reduce the circuit complexity, we propose
an efficient algorithm to minimize the number of RF switches while
maintaining the orthogonality among different pattern coders.

\textit{Finally}, we evaluate the spectral efficiency and energy efficiency
of the MIMO system with antenna coding using the pixel antennas. The
results show that the proposed technique using pixel antennas can
enhance the spectral efficiency of $4\times4$ MIMO by up to 12 bits/s/Hz
or equivalently reduces the required transmit power by up to 90\%
when compared to conventional MIMO system with fixed antenna configuration,
demonstrating the effectiveness of the proposed technique in exploiting
the spatial multiplexing to enhancing the wireless communication system.

\textit{Organization}: Section II provides the model of the pixel
antenna and introduces the antenna coding technique. Section III introduces
the MIMO system with antenna coding and derives the spectral efficiency
expressions of the system. Section IV analyzes the radiation pattern
and EADoF of the pixel antenna and proposes optimization algorithms
for antenna coding design. In Section V, we evaluate the spectral
efficiency and energy efficiency of the MIMO system to demonstrate
the effectiveness of antenna coding. Section VI concludes the work.

\textit{Notation}: Bold lower and upper case letters denote vectors
and matrices, respectively. Letters not in bold font represent scalars.
$\left|a\right|$ refers to the modulus of a complex scalar $a$.
$\left[\mathrm{\mathbf{a}}\right]_{i}$ and $\left\Vert \mathbf{a}\right\Vert $
refer to the $i$th entry and $l_{2}-$norm of vector $\mathrm{\mathbf{a}}$.
$\mathrm{\mathbf{A}}^{T}$, $\mathrm{\mathbf{A}}^{H}$, $\left[\mathrm{\mathbf{A}}\right]_{i,j}$,
and $\left|\mathrm{\mathbf{A}}\right|$ refer to the transpose, conjugate
transpose, $\left(i,j\right)$th entry, and determinant of a matrix
$\mathrm{\mathbf{A}}$, respectively. $\mathbb{R}$ and $\mathbb{C}$
denote the real and complex number sets, respectively. $\mathcal{A}\setminus\mathcal{B}$
denotes the set difference between set $\mathcal{A}$ and $\mathcal{B}$.
$\mathcal{CN}(\mu,\sigma^{2})$ denotes complex Gaussian distribution
with mean $\mu$ and variance $\sigma^{2}$. $\mathrm{prob}\left(x\right)$,
$\mathrm{H}\left(x\right)$ and $\mathrm{E}\left[x\right]$ refer
to the probability, entropy and expectation of a variable $x$. $\mathbf{I}_{M}$
denotes a $M\times M$ identity matrix. diag$\left(a_{1},...,a_{N}\right)$
refers to a diagonal matrix with diagonal elements being $a_{1},...,a_{N}$
and blkdiag$\left(\mathrm{\mathbf{A}}_{1},...,\mathrm{\mathbf{A}}_{N}\right)$
refers to a block diagonal matrix with diagonal matrices being $\mathrm{\mathbf{A}}_{1},...,\mathrm{\mathbf{A}}_{N}$.

\section{Pixel Antennas \label{sec:Pixel}}

\begin{figure}[t]
\begin{centering}
\includegraphics[width=8cm]{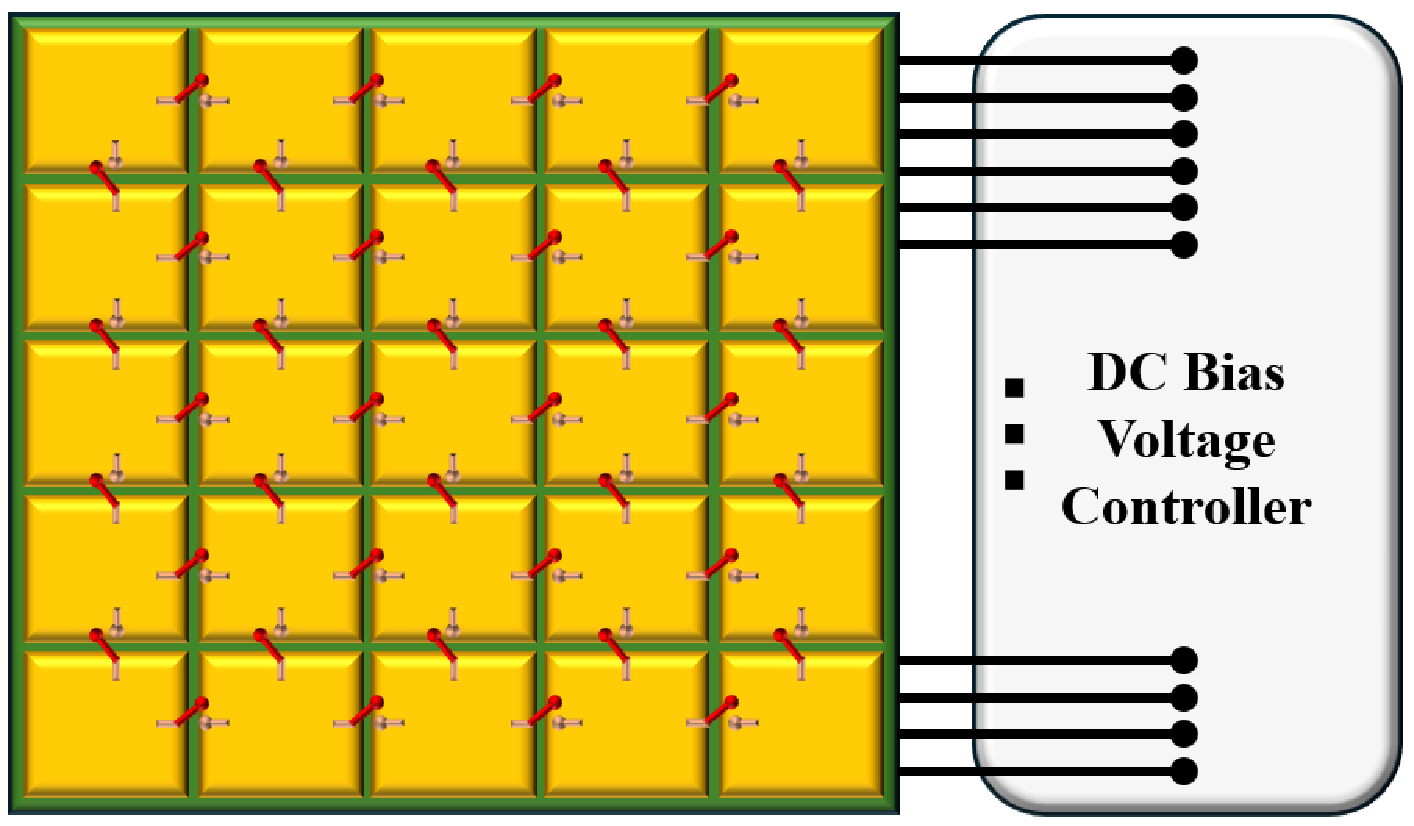}
\par\end{centering}
\caption{An illustrative example of a $5\times5$ pixel antenna. \label{fig:Pixel Antenna}}
\end{figure}

As illustrated in Fig. \ref{fig:Pixel Antenna}, a pixel antenna is
constructed by a set of individual sub-wavelength elements denoted
as pixels. This is a generalized framework for the reconfigurable
antenna since the antenna radiating structure of arbitrary shape and
size can be discretized into a grid of pixels. Adjacent pixels can
be connected or disconnected by using RF switches. Through adjusting
the direct current (DC) bias voltage to control the switch states,
the pixel antenna can be flexibly reconfigured, so that the surface
current and radiation pattern of the pixel antenna can be altered
to adapt to the channel.

\begin{figure}[t]
\begin{centering}
\includegraphics[width=8cm]{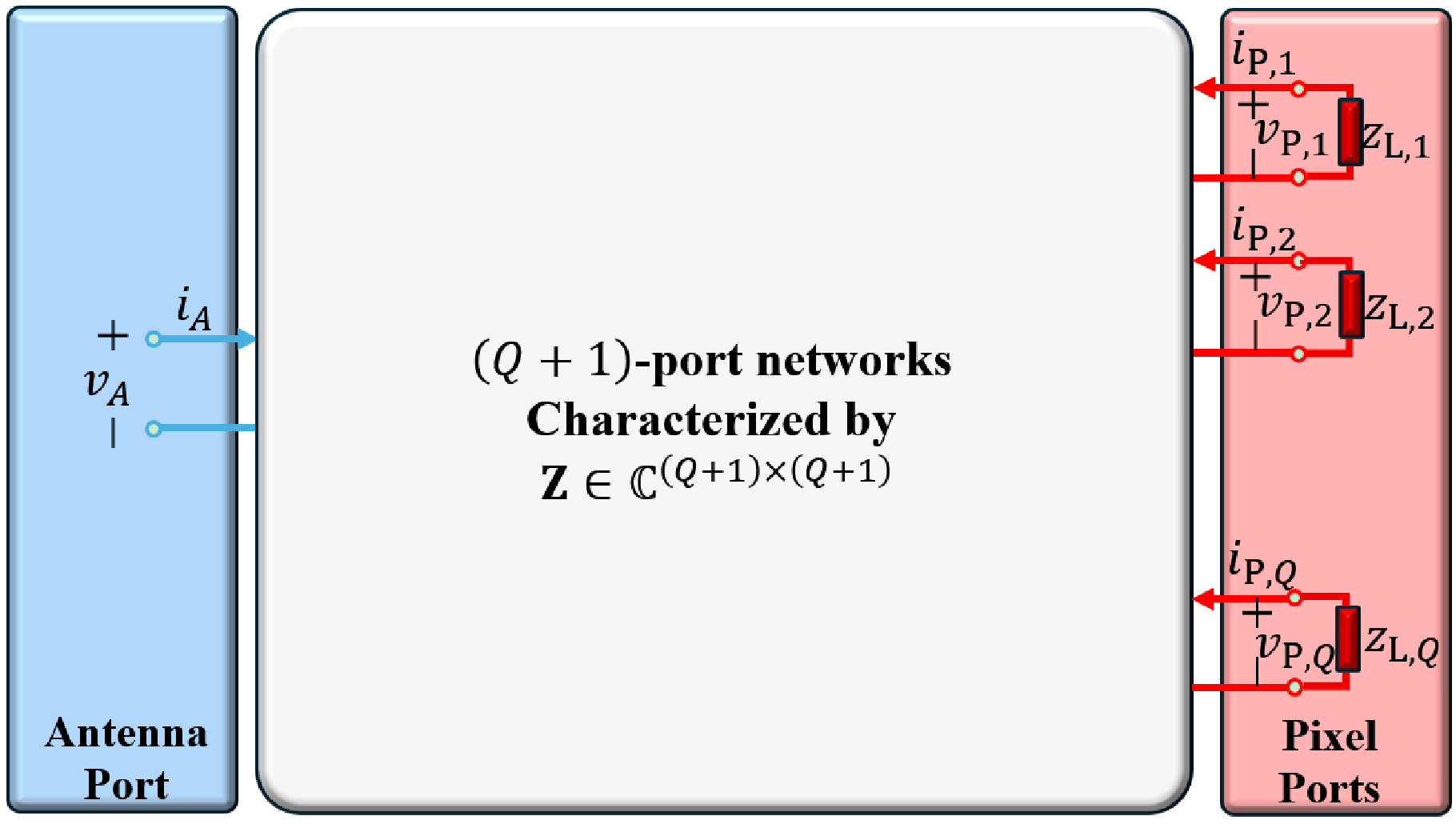}
\par\end{centering}
\caption{Equivalent $\left(Q+1\right)$-port network circuit model for the
pixel antenna with a single antenna port and $Q$ pixel ports. \label{fig:Equivalent}}
\end{figure}

To systematically analyze the pixel antenna, we can formulate the
equivalent circuit model as shown in Fig. \ref{fig:Equivalent} by
using multiport network theory. Specifically, for a pixel antenna
embedded with $Q$ switches, a $\left(Q+1\right)$-port network is
established where one antenna port is used to excite the pixel antenna
and $Q$ pixel ports are placed across adjacent pixels. The $\left(Q+1\right)$-port
network can be characterized by an impedance matrix $\mathbf{Z}\in\mathbb{C}^{\left(Q+1\right)\times\left(Q+1\right)}$
given by
\begin{equation}
\mathbf{Z}=\left[\begin{array}{ll}
z_{\mathrm{AA}} & \mathbf{z}_{\mathrm{AP}}\\
\mathbf{z}_{\mathrm{PA}} & \mathbf{Z}_{\mathrm{PP}}
\end{array}\right],\label{eq:Z}
\end{equation}
where $z_{\mathrm{AA}}\in\mathbb{C}$ is the self impedance of the
antenna port, $\mathbf{Z}_{\mathrm{PP}}\in\mathbb{C}^{Q\times Q}$
is the impedance sub-matrix of the $Q$ pixel ports, $\mathbf{z}_{\mathrm{AP}}\in\mathbb{C}^{1\times Q}$
and $\mathbf{z}_{\mathrm{PA}}\in\mathbb{C}^{Q\times1}$ are the mutual
impedance between the single antenna port and the $Q$ pixel ports
with $\mathbf{z}_{\mathrm{AP}}=\mathbf{z}_{\mathrm{PA}}^{T}$. The
antenna port is connected to an RF chain which provides feeding voltage
source, while the $q$th pixel port is connected to load impedance
$z_{\mathrm{L},q}$ which models the RF switch. Each RF switch has
two states, including switch on and off states, which can be denoted
as a binary variable $b_{q}\in\left\{ 0,1\right\} $ for $q=1,2,...,Q$,
and the corresponding load impedance can be either open- or short-circuit,
written as 
\begin{equation}
z_{\mathrm{L},q}=\begin{cases}
\infty, & b_{q}=1,\:\textrm{i.e. switch\:off},\\
0, & b_{q}=0,\:\textrm{i.e. switch\:on}.
\end{cases}\label{eq:Switch}
\end{equation}
Numerically we can use a very large value $z_{\mathrm{oc}}$ to approximate
the open-circuit load impedance $\infty$. Namely, a very large load
impedance can make adjacent pixels open-circuited which are equivalent
to being switched off, while zero load impedance can make adjacent
pixels short-circuited which are equivalent to being switched on.
We collect $b_{q}$, $\forall q$ into a binary vector $\mathbf{b}=\left[b_{1},b_{2},....,b_{Q}\right]^{T}\in\mathbb{\mathbb{R}}^{Q\times1}$
which can be coded to control the pixel antenna configuration and
this is referred to as an antenna coder. Accordingly, we can write
the load impedance for all $Q$ pixel ports as a diagonal matrix coded
by $\mathbf{b}$, that is $\mathbf{Z}_{\mathrm{L}}\left(\mathbf{b}\right)=\mathrm{diag}\left(z_{\mathrm{L},1},z_{\mathrm{L},2},...,z_{\mathrm{L},Q}\right)=z_{\mathrm{oc}}\mathrm{diag}\left(b_{1},b_{2},....,b_{Q}\right)\in\mathbb{C}^{Q\times Q}$.

Using \eqref{eq:Z} and \eqref{eq:Switch}, the voltage and current
in the $\left(Q+1\right)$-port network are related by \cite{5446312}
\begin{equation}
\left[\begin{array}{c}
v_{\mathrm{A}}\\
\mathbf{v}_{\mathrm{P}}
\end{array}\right]=\left[\begin{array}{ll}
z_{\mathrm{AA}} & \mathbf{z}_{\mathrm{AP}}\\
\mathbf{z}_{\mathrm{PA}} & \mathbf{Z}_{\mathrm{PP}}
\end{array}\right]\left[\begin{array}{c}
i_{\mathrm{A}}\\
\mathbf{i}_{\mathrm{P}}
\end{array}\right],\label{eq:V=00003DZI}
\end{equation}
where $v_{\mathrm{A}}\in\mathbb{C}$ and $\mathbf{v}_{\mathrm{P}}\in\mathbb{C}^{Q\times1}$
are the voltages across the antenna and pixel ports, respectively,
$i_{\mathrm{A}}\in\mathbb{C}$ and $\mathbf{i}_{\mathrm{P}}\in\mathbb{C}^{Q\times1}$
are the currents through the antenna and pixel ports, respectively.
In addition, the voltage and current at the pixel ports are related
by
\begin{equation}
\mathbf{v}_{\mathrm{P}}=-\mathbf{Z}_{\mathrm{L}}\left(\mathbf{b}\right)\mathbf{i}_{\mathrm{P}}.\label{eq:V=00003DZI-2}
\end{equation}
Substituting \eqref{eq:V=00003DZI-2} into \eqref{eq:V=00003DZI},
we can obtain the relationship between the current on the antenna
port and pixel ports as
\begin{equation}
\mathbf{i}_{\mathrm{P}}\left(\mathbf{b}\right)=-\left(\mathbf{Z}_{\mathrm{PP}}+\mathbf{Z}_{\mathrm{L}}\left(\mathbf{b}\right)\right)^{-1}\mathbf{z}_{\mathrm{PA}}i_{\mathrm{A}}.\label{eq:current relation}
\end{equation}
We collect the currents through all ports of the pixel antenna as
\begin{equation}
\mathbf{i}\left(\mathbf{b}\right)=\left[\begin{array}{c}
i_{\mathrm{A}}\\
\mathbf{i}_{\mathrm{P}}\left(\mathbf{b}\right)
\end{array}\right]=\left[\begin{array}{c}
1\\
-\left(\mathbf{Z}_{\mathrm{PP}}+\mathbf{Z}_{\mathrm{L}}\left(\mathbf{b}\right)\right)^{-1}\mathbf{z}_{\mathrm{PA}}
\end{array}\right]i_{\mathrm{A}}.\label{eq:current coder}
\end{equation}
so that pixel antenna current $\mathbf{i}$ can be coded by the antenna
coder $\mathbf{b}$.

Our aim is to perform antenna coding based on pixel antennas for MIMO
communications within the framework of a beamspace channel representation.
To that end, the radiation pattern of the pixel antenna, denoted as
$\mathbf{e}\left(\mathbf{b}\right)=\left[\mathbf{e}_{\theta},\mathbf{e}_{\phi}\right]^{T}\in\mathbb{C}^{2K\times1}$
where $\mathbf{e}_{\theta}$ and $\mathbf{e}_{\phi}$ are the elevation
and azimuth polarization components sampled at $K$ spatial angles,
is represented as a function of antenna coder $\mathbf{b}$, which
is given by
\begin{align}
\mathbf{e}\left(\mathbf{b}\right) & =\mathbf{e}_{\mathrm{A}}i_{\mathrm{A}}+\mathbf{E}_{\mathrm{P}}\mathbf{i}_{\mathrm{P}}\left(\mathbf{b}\right)=\mathbf{E}_{\mathrm{oc}}\mathbf{i}\left(\mathbf{b}\right),\label{eq:Pattern}
\end{align}
where $\mathbf{e}_{\mathrm{A}}=[\mathbf{e}_{\mathrm{A},\theta},\mathbf{e}_{\mathrm{A},\phi}]^{T}\in\mathbb{C}^{2K\times1}$
is the open-circuit radiation pattern\footnote{The open-circuit radiation pattern refers to the radiation pattern
excited by unit current with all the other ports open-circuit.} of the antenna port with $\mathbf{e}_{\mathrm{A},\theta}$ and $\mathbf{e}_{\mathrm{A},\phi}$
being the two polarization components and $\mathbf{E}_{\mathrm{P}}=[\mathbf{e}_{\mathrm{P},1},\ldots,\mathbf{e}_{\mathrm{P},Q}]\in\mathbb{C}^{2K\times Q}$
collects the open-circuit radiation pattern of pixel ports, $\mathbf{e}_{\mathrm{P},q}=[\mathbf{e}_{\mathrm{P},q,\theta},\mathbf{e}_{\mathrm{P},q,\phi}]^{T}\in\mathbb{C}^{2K\times1}$,
for $q=1,2,...,Q$, into a matrix with $\mathbf{e}_{\mathrm{P},q,\theta}$
and $\mathbf{e}_{\mathrm{P},q,\phi}$ being the two polarization components.
$\mathbf{E}_{\mathrm{oc}}=\left[\mathbf{e}_{\mathrm{A}},\mathbf{E}_{\mathrm{P}}\right]\in\mathbb{C}^{2K\times\left(Q+1\right)}$
collects the open-circuit radiation patterns of all $Q+1$ ports of
the pixel antenna, referred to as the open-circuit radiation pattern
matrix. We can observe that the radiation pattern of the pixel antenna
$\mathbf{e}\left(\mathbf{b}\right)$ consists of a perturbation term
that can be coded by the antenna coder $\mathbf{b}$, which will be
further analyzed in Section IV. Therefore, antenna coding enables
control of the radiation pattern of the pixel antenna for MIMO communication
within the framework of the beamspace channel representation.

The open-circuit radiation pattern matrix of the pixel antenna can
be decomposed by using singular value decomposition as
\begin{equation}
\mathbf{E}_{\mathrm{oc}}=\mathbf{U}\mathbf{\Sigma}\mathbf{V}^{H},\label{eq:SVD}
\end{equation}
where $\mathbf{U}\in\mathbb{C}^{2K\times R}$ and $\mathbf{V}\in\mathbb{C}^{\left(Q+1\right)\times R}$
are semi-unitary matrices satisfying $\mathbf{U}^{H}\mathbf{U}=\mathbf{I}_{R}$
and $\mathbf{V}^{H}\mathbf{V}=\mathbf{I}_{R}$ with $R=\mathrm{rank}\left(\mathbf{E}_{\mathrm{oc}}\right)$
being the rank of matrix $\mathbf{E}_{\mathrm{oc}}$ and can be considered
as the number of orthonormal basis radiation patterns that the pixel
antenna can provide, i.e. the effective aerial DoF (EADoF). $\mathbf{\Sigma}=\mathrm{diag}\left(\sigma_{1},\sigma_{2},...,\sigma_{R}\right)\in\mathbb{R}^{R\times R}$
is a diagonal matrix with diagonal entries being singular values.
In essence, each column in $\mathbf{U}$ can be regarded as one orthonormal
basis radiation pattern. Taking \eqref{eq:SVD} into \eqref{eq:Pattern},
we have
\begin{align}
\mathbf{e}\left(\mathbf{b}\right) & =\mathbf{U}\mathbf{\Sigma}\mathbf{V}^{H}\mathbf{i}\left(\mathbf{b}\right)=\mathbf{U}\mathbf{w}\left(\mathbf{b}\right),\label{eq:pattern coder}
\end{align}
which implies that any radiation pattern of the pixel antenna can
be decomposed into the $R$ orthonormal basis radiation patterns,
and $\mathbf{w}\left(\mathbf{b}\right)\triangleq\mathbf{\Sigma}\mathbf{V}^{H}\mathbf{i}\left(\mathbf{b}\right)\in\mathbb{C}^{R\times1}$
characterizes how to linearly code the $R$ orthonormal basis radiation
patterns to construct the radiation pattern of the pixel antenna,
and is referred to here as a \textit{pattern coder}. In the following,
we will show how to design the antenna coder $\mathbf{b}$ to modulate
the pattern coder $\mathbf{w}\left(\mathbf{b}\right)$, so as to enhance
the spectral efficiency of MIMO communication systems.

\section{MIMO System with Antenna Coding\label{sec:System}}

In this section, we firstly provide the MIMO system model using the
beamspace channel representation with antenna coding. Then, we propose
enhancing the spectral efficiency of MIMO system via the antenna coding
technique based on the pixel antenna. This is in contrast to a conventional
MIMO system with a fixed antenna configuration.

\subsection{Beamspace Channel Representation with Antenna Coding}

\begin{figure*}[t]
\centering{}\includegraphics[width=7.5cm]{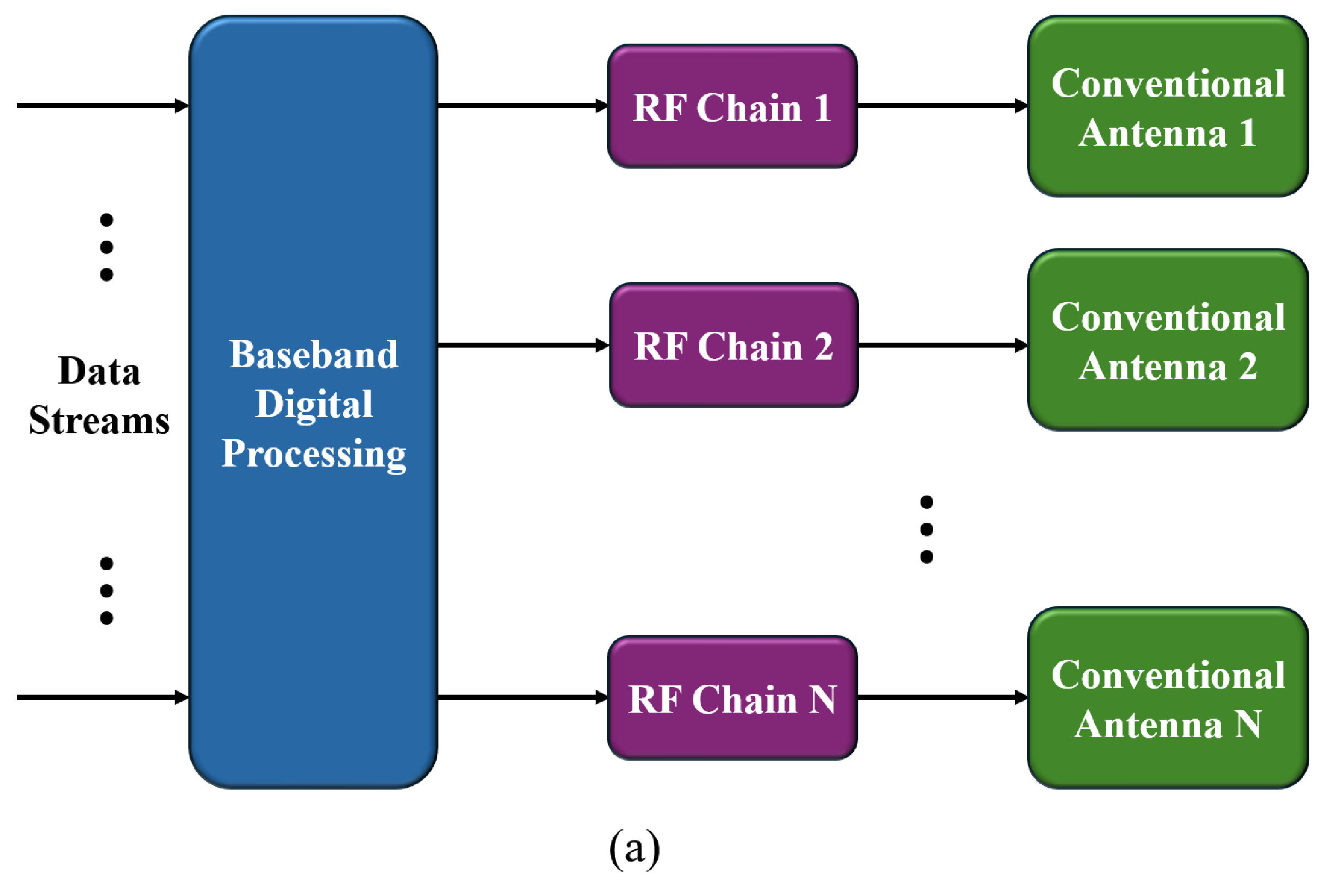}$\qquad\qquad$\includegraphics[width=7.5cm]{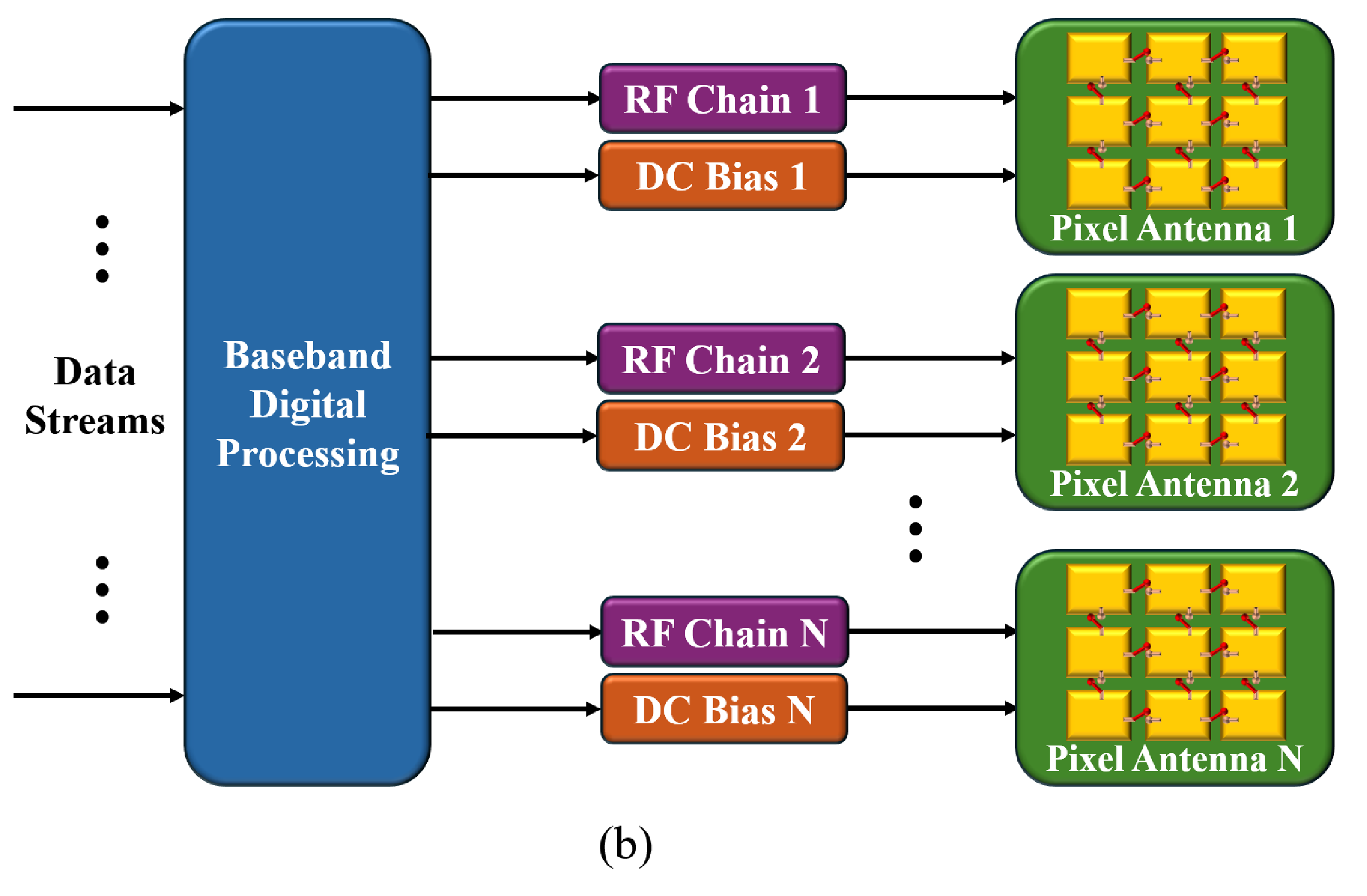}\caption{Schematic of the transmitter for (a) conventional MIMO system with
fixed antenna configuration and (b) MIMO system with pixel antennas.
\label{fig:MIMO Comparison}}
\end{figure*}

Consider a MIMO system with $N$ transmit antennas and $M$ receive
antennas where each antenna is connected to an RF chain. Antennas
at both the transmitter and receiver are spatially isolated by at
least half wavelength so that we can assume no mutual coupling.

We start from the model of a conventional MIMO system with a fixed
antenna configuration, as shown in Fig. \ref{fig:MIMO Comparison}(a),
which can be written as 
\begin{equation}
\mathbf{y}=\mathbf{H}\mathbf{s}+\mathbf{n},\label{eq:system}
\end{equation}
where $\mathbf{y}\in\mathbb{C}^{M\times1}$ is the received signal,
$\mathbf{H}\in\mathbb{C}^{M\times N}$ is the channel matrix, $\mathbf{s}=\left[s_{1},s_{2},...,s_{N}\right]^{T}\in\mathbb{C}^{N\times1}$
is the transmit symbol with $s_{n}$ for $n=1,2,...,N$ being the
digitally modulated symbol output by the $n$th RF chain at the transmitter,
and $\mathbf{n}\sim\mathcal{CN}\left(0,N_{0}\mathbf{I}_{N}\right)$
is the additive white Gaussian noise with power of $N_{0}$.

Different from the conventional MIMO system with fixed antenna configuration,
the MIMO system with pixel antennas can flexibly reconfigure the radiation
patterns of the antennas to bring extra DoF for system enhancement.
To characterize the role of radiation pattern reconfigurability of
the pixel antenna in MIMO system, we consider the beamspace channel
representation. In the beamspace domain, transmit symbols are mapped
into a set of orthonormal basis radiation patterns instead of into
independent spatially isolated antennas as in conventional MIMO systems
\cite{kalis2008novel}\nocite{alrabadi2009universal}\nocite{alrabadi2010spatial}\nocite{alrabadi2011mimo}-\cite{barousis2011beamspace}.
Accordingly, we can rewrite the channel matrix $\mathbf{H}$ in the
beamspace domain with $K$ sampled spatial angles as
\begin{equation}
\mathbf{H}=\mathbf{F}^{T}\mathbf{H}_{\mathrm{V}}\mathbf{E},\label{eq:BS H}
\end{equation}
where $\mathbf{E}=\left[\mathbf{e}_{1},\mathbf{e}_{2},...,\mathbf{e}_{N}\right]\in\mathbb{C}^{2K\times N}$
consists of the radiation patterns of the $N$ transmit antennas $\mathbf{e}_{n}=[\mathbf{e}_{\theta,n},\mathbf{e}_{\phi,n}]^{T}\in\mathbb{C}^{2K\times1}$
for $n=1,2,...,N$ with normalization $\left\Vert \mathbf{e}_{n}\right\Vert ^{2}=1$
and $\mathbf{F}=\left[\mathbf{f}_{1},\mathbf{f}_{2},\ldots,\mathbf{f}_{M}\right]\in\mathbb{C}^{2K\times M}$
consists of the radiation patterns of the $M$ receive antennas $\mathbf{f}_{m}=[\mathbf{f}_{\theta,m},\mathbf{f}_{\phi,m}]^{T}\in\mathbb{C}^{2K\times1}$
for $m=1,2,...,M$ with normalization $\left\Vert \mathbf{f}_{m}\right\Vert ^{2}=1$.
$\mathbf{H}_{\mathrm{V}}$ is the virtual channel matrix given by
\begin{equation}
\mathbf{H}_{\mathrm{V}}=\left[\begin{array}{cc}
\mathbf{H}_{\mathrm{V},\theta\theta} & \mathbf{H}_{\mathrm{V},\theta\phi}\\
\mathbf{H}_{\mathrm{V},\phi\theta} & \mathbf{H}_{\mathrm{V},\phi\phi}
\end{array}\right],\label{eq:Hv}
\end{equation}
where $\mathbf{H}_{\mathrm{V},\theta\theta}$, $\mathbf{H}_{\mathrm{V},\theta\phi}$,
$\mathbf{H}_{\mathrm{V},\phi\theta}$, and $\mathbf{H}_{\mathrm{V},\phi\phi}\in\mathbb{C}^{K\times K}$
are the virtual channel matrices for the elevation and azimuth polarizations,
respectively \cite{sayeed2002deconstructing}-\nocite{maliatsos2013modifications}\cite{shafi2006polarized},
with each entry being the channel gain from an angle of departure
to an angle of arrival among $K$ samples. We also assume a rich scattering
environment so that entries in $\mathbf{H}_{\mathrm{V}}$ follow the
i.i.d. complex Gaussian distribution $\mathcal{CN}\left(0,1\right)$.

Replacing each of the transmit antennas in the conventional MIMO system
with a pixel antenna, as illustrated in Fig. \ref{fig:MIMO Comparison}
(b), the radiation patterns $\mathbf{E}$ in \eqref{eq:BS H} are
no longer constant, and instead can be coded by the antenna coders
written as $\mathbf{E}\left(\mathbf{B}\right)=\left[\mathbf{e}_{1}\left(\mathbf{b}_{1}\right),\mathbf{e}_{2}\left(\mathbf{b}_{2}\right),...,\mathbf{e}_{N}\left(\mathbf{b}_{N}\right)\right]$
where $\mathbf{B}=\left[\mathbf{b}_{1},\mathbf{b}_{2},...,\mathbf{b}_{N}\right]\in\mathbb{R}^{Q\times N}$
collects the antenna coders for all the $N$ pixel antennas, i.e.
$\mathbf{b}_{n}$, $\forall n$. Thus, the beamspace channel matrix
$\mathbf{H}$ in \eqref{eq:BS H} can be rewritten as
\begin{equation}
\mathbf{H}\left(\mathbf{B}\right)=\mathbf{F}^{T}\mathbf{H}_{\mathrm{V}}\mathrm{\mathbf{E}}\left(\mathbf{B}\right).\label{eq:H Current}
\end{equation}
In addition, the radiation pattern of the $n$th pixel antenna $\mathbf{e}_{n}\left(\mathbf{b}_{n}\right)$,
coded by the antenna coder $\mathbf{b}_{n}$, can be found by
\begin{equation}
\mathbf{e}_{n}\left(\mathbf{b}_{n}\right)=\mathbf{E}_{\mathrm{oc},n}\mathbf{i}_{n}\left(\mathbf{b}_{n}\right)=\mathbf{U}_{n}\mathbf{w}_{n}\left(\mathbf{b}_{n}\right),\label{eq: enbn =00003D unwn}
\end{equation}
where $\mathbf{E}_{\mathrm{oc},n}$ and $\mathbf{i}_{n}\left(\mathbf{b}_{n}\right)$
are the open-circuit radiation pattern matrix and the coded current
for the $n$th pixel antenna, respectively. $\mathbf{U}_{n}$ collects
the orthonormal basis radiation patterns of the $n$th pixel antenna
obtained by the singular value decomposition, i.e. $\mathbf{E}_{\mathrm{oc},n}=\mathbf{U}_{n}\mathbf{\Sigma}_{n}\mathbf{V}_{n}^{H}$,
so that we have 
\begin{equation}
\mathbf{w}_{n}\left(\mathbf{b}_{n}\right)=\mathbf{\Sigma}_{n}\mathbf{V}_{n}^{H}\mathbf{i}_{n}\left(\mathbf{b}_{n}\right),
\end{equation}
which is the pattern coder for the $n$th pixel antenna. Substituting
\eqref{eq: enbn =00003D unwn} into \eqref{eq:H Current}, we can
rewrite \eqref{eq:H Current} as
\begin{equation}
\mathbf{H}=\mathbf{F}^{T}\mathbf{H}_{\mathrm{V}}\mathbf{U}_{\mathrm{BS}}\mathbf{W}\left(\mathbf{B}\right)\label{eq:H BS}
\end{equation}
where $\mathrm{\mathbf{U}_{BS}}=\left[\mathbf{U}_{1},\mathbf{U}_{2},...,\mathbf{U}_{N}\right]\in\mathbb{C}^{2K\times NR}$
represents the matrix collecting all the orthonormal basis radiation
patterns to form the beamspace and $\mathbf{W}\left(\mathbf{\mathbf{B}}\right)=\mathrm{blkdiag}\left(\mathbf{w}_{1}\left(\mathbf{b}_{1}\right),\mathbf{w}_{2}\left(\mathbf{b}_{2}\right),...,\mathbf{w}_{N}\left(\mathbf{b}_{N}\right)\right)\in\mathbb{\mathbb{C}}^{NR\times N}$
is a block diagonal matrix with diagonal blocks being pattern coders
$\mathbf{w}_{n}\left(\mathbf{b}_{n}\right)$, $n=1,2,...,N$. Substituting
\eqref{eq:H BS} into \eqref{eq:system}, the overall system model
is expressed as
\begin{align}
\mathbf{y} & =\mathbf{F}^{T}\mathbf{H}_{\mathrm{V}}\mathbf{U}_{\mathrm{BS}}\mathbf{W}\left(\mathbf{B}\right)\mathbf{s}+\mathbf{n}\nonumber \\
 & =\mathbf{F}^{T}\mathbf{H}_{\mathrm{V}}\mathbf{U}_{\mathrm{BS}}\mathbf{x}\left(\mathbf{B},\mathbf{s}\right)+\mathbf{n}=\mathbf{H}_{\mathrm{BS}}\mathbf{x}\left(\mathbf{B},\mathbf{s}\right)+\mathbf{n},\label{eq:BS system}
\end{align}
where $\mathbf{x}\left(\mathbf{B},\mathbf{s}\right)=\mathbf{W}\left(\mathbf{B}\right)\mathbf{s}\in\mathbb{C}^{NR\times1}$
is the equivalent beamspace transmit signal and $\mathbf{H}_{\mathrm{BS}}=\mathbf{F}^{H}\mathbf{H}_{\mathrm{V}}\mathbf{U}_{\mathrm{BS}}$
is the equivalent beamspace channel matrix. It should be noted that
as $\mathbf{F}$ and $\mathbf{U}_{\mathrm{BS}}$ consist of orthonormal
basis radiation patterns, we can assume that each entry in $\mathbf{H}_{\mathrm{BS}}$
follows the i.i.d. complex Gaussian distribution $\mathcal{CN}\left(0,1\right)$,
which is equivalent to the conventional MIMO channel with Rayleigh
fading.

\subsection{Spectral Efficiency Enhancement via Antenna Coding}

\begin{figure}[t]
\begin{centering}
\includegraphics[width=8.5cm]{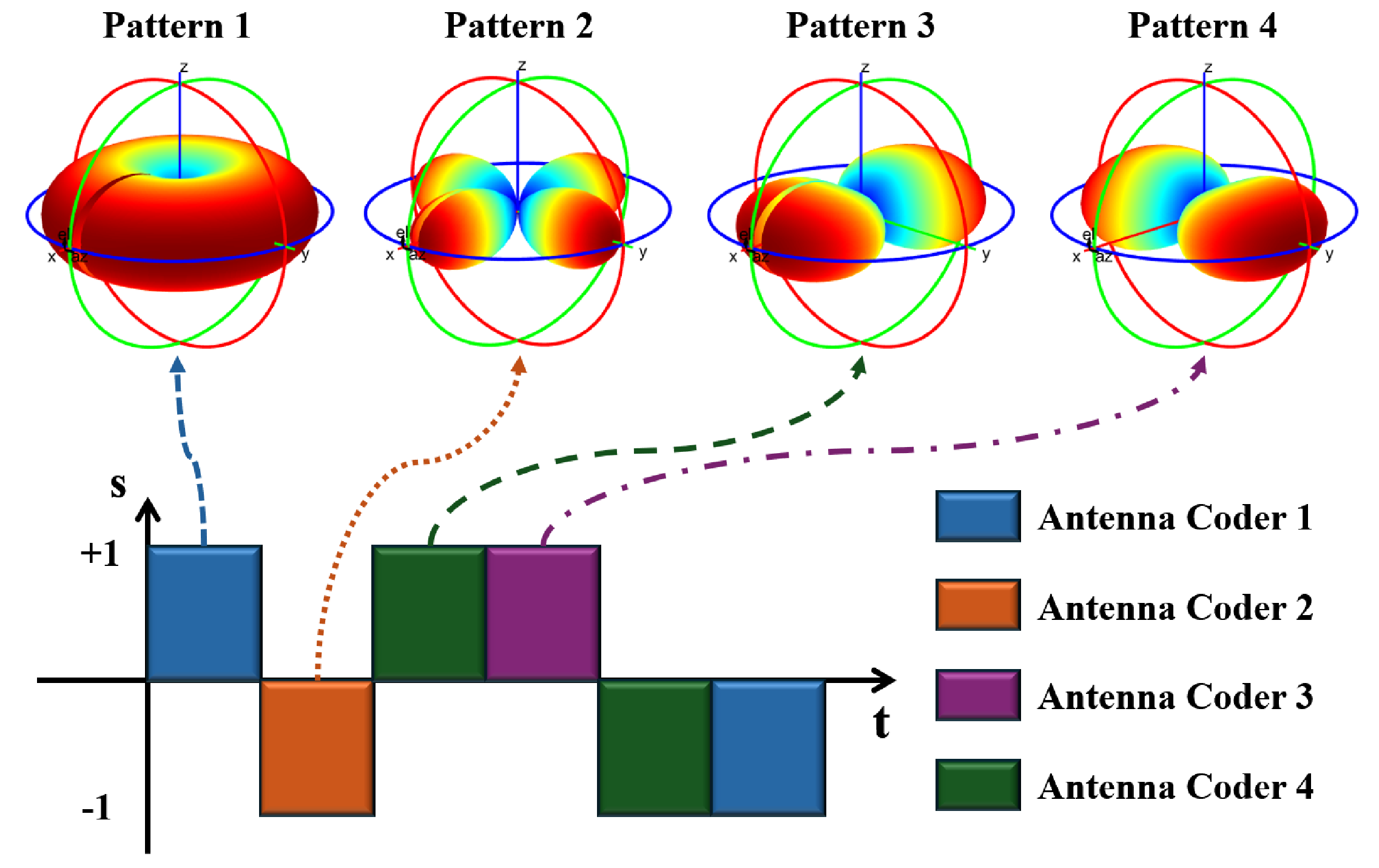}
\par\end{centering}
\caption{An illustration of simultaneously transmitting digital modulated symbol,
e.g. binary phase shift keying (BPSK), and performing antenna coding
to select radiation pattern by a pixel antenna to enhance the spectral
efficiency. \label{fig:TimePattern}}
\end{figure}
We can notice that $\mathbf{B}$ and $\mathbf{s}$ in $\mathbf{x}\left(\mathbf{B},\mathbf{s}\right)$
are mutually independent from \eqref{eq:BS system}. As illustrated
in Fig. \ref{fig:TimePattern}, each pixel antenna can simultaneously
transmit the digital modulated symbol $s$ and perform antenna coding
to select radiation pattern. Therefore, additional information bits
can be modulated and transmitted by the antenna coder matrix $\mathbf{B}$
through controlling DC bias voltages on the RF switches of pixel antennas
as shown in Fig. \ref{fig:MIMO Comparison}(b). Specifically, for
the transmit pixel antenna, we can define a codebook consisting of
$P$ different antenna coders written as 
\begin{equation}
\mathcal{B}=\left\{ \mathbf{b}_{\mathrm{C},1},\mathbf{b}_{\mathrm{C},2},...,\mathbf{b}_{\mathrm{C},P}\right\} ,\label{eq:coderbook}
\end{equation}
where the $p$th antenna coder $\mathbf{b}_{\mathrm{C},p}$ is associated
to the $p$th pattern coder $\mathbf{w}\left(\mathbf{b}_{\mathrm{C},p}\right)$.
The indices of antenna coders in $\mathcal{B}$ can be mapped into
bits and modulated to digital symbols. Thus, by selecting one of the
$P$ antenna coders for each transmit pixel antenna, additional symbols
can be modulated on antenna coders and the resulting radiation patterns
can be transmitted so that the spatial multiplexing can be exploited
to enhance spectral efficiency of MIMO systems. As a demonstration
for the effectiveness of the proposed antenna coding technique in
exploiting the spatial multiplexing, for each transmit pixel antenna,
we consider the antenna coding design by selecting a radiation pattern,
in each symbol period, among $P$ basis radiation patterns $\left(P\leqslant R\right)$.
That is modulating the transmit symbol among the $P$ basis radiation
patterns. Therefore, the corresponding $P$ pattern coders are given
by
\begin{equation}
\mathbf{w}\left(\mathbf{b}_{\mathrm{C},p}\right)=\underset{\overset{\shortdownarrow\:\;\:\;}{\textrm{the }p\textrm{th entry}}}{\left[0,..,0,1,0,...,0\right]^{T}},\forall p,\label{eq:pattern coder SSK}
\end{equation}
where the $p$th entry is unity with the other entries being zero
for $p=1,2,...,P$. In other words, for the $n$th transmit pixel
antenna, its antenna coder $\mathbf{b}_{n}$ can be modulated, as
illustrated in Fig. \ref{fig:TimePattern}, by selecting among the
antenna coders $\mathbf{b}_{C,p}$, $\forall p$, that is exciting
among the $P$ orthonormal basis radiation patterns in $\mathbf{U}_{n}$.
It should be noted that this technique can be extended to other forms
of antenna coders by setting the pattern coder $\mathbf{w}\left(\mathbf{b}_{\mathrm{C},p}\right)$
to different values.

Based on the system model \eqref{eq:BS system} and modulated pattern
coder \eqref{eq:pattern coder SSK}, we next analyze how spectral
efficiency of MIMO system is enhanced with the proposed antenna coding.
With $N$ transmit pixel antennas and $P$ antenna coders for each
pixel antenna, there are in total $P^{N}$ combinations for the antenna
coder matrix $\mathbf{B}$ to transmit additional information bits
for spectral efficiency enhancement. The mutual information (MI) of
this MIMO system $\mathrm{I}\left(\mathbf{x}\left(\mathbf{B},\mathbf{s}\right);\mathbf{y}|\mathbf{H}_{\mathrm{BS}}\right)$
can be decomposed onto the MI of the transmit modulated symbol $\mathbf{s}$
and the MI of the antenna coding $\mathbf{W}\left(\mathbf{B}\right)$
\cite{6877673} as
\begin{align}
\mathrm{I}\left(\mathbf{x}\left(\mathbf{B},\mathbf{s}\right);\mathbf{y}|\mathbf{H}_{\mathrm{BS}}\right) & =\mathrm{I}\left(\mathbf{s};\mathbf{y}|\mathbf{W}\left(\mathbf{B}\right),\mathbf{H}_{\mathrm{BS}}\right)\nonumber \\
 & \:\quad+\mathrm{I}\left(\mathbf{W}\left(\mathbf{B}\right);\mathbf{y}|\mathbf{H}_{\mathrm{BS}}\right).\label{eq:mutual information}
\end{align}
It can be straightforwardly observed that the first term can be transformed
onto MI of a conventional MIMO as
\begin{align}
 & \mathrm{I}\left(\mathbf{s};\mathbf{y}|\mathbf{W}\left(\mathbf{B}\right),\mathbf{H}_{\mathrm{BS}}\right)=\mathrm{I}\left(\mathbf{s};\mathbf{y}|\mathbf{H}_{\mathrm{BS}}\mathbf{W}\left(\mathbf{B}\right)\right)\nonumber \\
 & =\mathrm{H}\left(\mathbf{y}|\mathbf{H}_{\mathrm{BS}}\mathbf{W}\left(\mathbf{B}\right)\right)-\mathrm{H}\left(\mathbf{y}|\mathbf{H}_{\mathrm{BS}}\mathbf{W}\left(\mathbf{B}\right),\mathbf{s}\right)\nonumber \\
 & =\mathrm{\log}_{2}\left|\mathbf{I}_{M}+\frac{\mathbf{H}_{\mathrm{BS}}\mathbf{W}\left(\mathbf{B}\right)\mathbf{S}\mathbf{W}^{H}\left(\mathbf{B}\right)\mathbf{H}_{\mathrm{BS}}^{H}}{N_{0}}\right|,\label{eq:MIMO MI}
\end{align}
where $\mathbf{S}=\mathrm{E}\left[\mathbf{s}\mathbf{s}^{H}\right]$
is the covariance matrix of transmit symbol $\mathbf{s}$. Then the
second term in \eqref{eq:mutual information} can be expressed as
\begin{equation}
\mathrm{I}\left(\mathbf{W}\left(\mathbf{B}\right);\mathbf{y}|\mathbf{H}_{\mathrm{BS}}\right)=\mathrm{H}\left(\mathbf{W}\left(\mathbf{B}\right)|\mathbf{H}_{\mathrm{BS}}\right)-\mathrm{H}\left(\mathbf{W}\left(\mathbf{B}\right)|\mathbf{H}_{\mathrm{BS}},\mathbf{y}\right),\label{eq:SSK MI}
\end{equation}
where $\mathrm{H}\left(\mathbf{W}\left(\mathbf{B}\right)|\mathbf{H}_{\mathrm{BS}}\right)=N\mathrm{\log}_{2}P$
since $\mathbf{W}\left(\mathbf{B}\right)$ is independent of beamspace
channel $\mathbf{H}_{\mathrm{BS}}$. $\mathrm{H}\left(\mathbf{W}\left(\mathbf{B}\right)|\mathbf{H}_{\mathrm{BS}},\mathbf{y}\right)$
can be further given by
\begin{align}
 & \mathrm{H}\left(\mathbf{W}\left(\mathbf{B}\right)|\mathbf{H}_{\mathrm{BS}},\mathbf{y}\right)\nonumber \\
 & =\mathrm{E}_{\mathbf{W}\left(\mathbf{B}\right),\mathbf{H}_{\mathrm{BS}},\mathbf{y}}\left[\mathrm{\mathrm{\log}_{2}}\textrm{prob}\left(\mathbf{W}\left(\mathbf{B}\right)|\mathbf{H}_{\mathrm{BS}},\mathbf{y}\right)^{-1}\right].\label{eq:SSK MI EXPECT}
\end{align}
Using Bayes theorem we have that
\begin{align}
 & \textrm{prob}\left(\mathbf{W}\left(\mathbf{B}\right)|\mathbf{H}_{\mathrm{BS}},\mathbf{y}\right)\nonumber \\
 & =\frac{\textrm{prob}\left(\mathbf{y}|\mathbf{H}_{\mathrm{BS}}\mathbf{W}\left(\mathbf{B}\right)\right)\textrm{prob}\left(\mathbf{W}\left(\mathbf{B}\right)|\mathbf{H}_{\mathrm{BS}}\right)}{\sum_{i=1}^{P^{N}}\textrm{prob}\left(\mathbf{y}|\mathbf{H}_{\mathrm{BS}}\mathbf{W}\left(\mathbf{B}_{i}\right)\right)\textrm{prob}\left(\mathbf{W}\left(\mathbf{B}_{i}\right)|\mathbf{H}_{\mathrm{BS}}\right)}\nonumber \\
 & =\frac{\textrm{prob}\left(\mathbf{y}|\mathbf{H}_{\mathrm{BS}}\mathbf{W}\left(\mathbf{B}\right)\right)}{\sum_{i=1}^{P^{N}}\textrm{prob}\left(\mathbf{y}|\mathbf{H}_{\mathrm{BS}}\mathbf{W}\left(\mathbf{B}_{i}\right)\right)}\nonumber \\
 & =\frac{\mathrm{E}_{\mathbf{s}}\left[\mathrm{exp}\left(-\frac{\left\Vert \mathbf{y}-\mathbf{H}_{\mathrm{BS}}\mathbf{W}\left(\mathbf{B}\right)\mathbf{s}\right\Vert ^{2}}{N_{0}}\right)\right]}{\sum_{i=1}^{P^{N}}\mathrm{E}_{\mathbf{s}}\left[\mathrm{exp}\left(-\frac{\left\Vert \mathbf{y}-\mathbf{H}_{\mathrm{BS}}\mathbf{W}\left(\mathbf{B}_{i}\right)\mathbf{s}\right\Vert ^{2}}{N_{0}}\right)\right]},\label{eq:SSK MI Results}
\end{align}
where $\mathbf{B}_{i}$ denotes the $i$th antenna coder matrix among
the $P^{N}$ combinations for the antenna coder matrix $\mathbf{B}$
and $\mathrm{prob}\left(\mathbf{W}\left(\mathbf{B}_{i}\right)|\mathbf{H}_{\mathrm{BS}}\right)=P^{-N}$
for $i=1,2,...,P^{N}$ because of the $P^{N}$ combinations $\mathbf{B}_{i}$,
$\forall i$ are randomly selected with equal probability. By taking
\eqref{eq:SSK MI Results} into \eqref{eq:SSK MI EXPECT} and taking
\eqref{eq:MIMO MI} and \eqref{eq:SSK MI} into \eqref{eq:mutual information},
we obtain the overall spectral efficiency of MIMO system with the
proposed antenna coding technique. It can be observed that the antenna
coding increases the spectral efficiency, which is reflected by the
second term in \eqref{eq:mutual information}, by transmitting additional
information associated with the radiation pattern selection of transmit
pixel antennas. Particularly, when the signal-to-noise ratio (SNR)
is large, the spectral efficiency can be approximately increased by
$N\mathrm{\log}_{2}P$. In practical applications, we can select $P$
as the positive integer power of two so that $\mathrm{\log}_{2}P$
is an integer.

\section{Pixel Antenna Analysis and Optimization \label{sec:Optimization}}

In this section, we analyze the radiation pattern of the pixel antenna
with antenna coding and the EADoF. In addition, we propose efficient
optimization algorithms for designing the antenna coders which achieve
the $P$ modulated pattern coders in \eqref{eq:pattern coder SSK}
with least number of RF switches.

\subsection{Analysis for Pixel Antenna Radiation Pattern}

It can be observed from \eqref{eq:current relation} that the antenna
coder $\mathbf{b}$ is contained in a diagonal matrix $\mathbf{Z}_{\mathrm{L}}$
while matrix inversion is also involved in the perturbation term.
This makes the analysis of the radiation pattern difficult. To simplify
the perturbation term, we firstly extract the diagonal entries in
$\mathbf{Z}_{\mathrm{PP}}$ as $\mathbf{Z}_{\mathrm{PP,D}}=\mathrm{diag}\left(\left[\mathbf{Z}_{\mathrm{PP}}\right]_{1,1},\left[\mathbf{Z}_{\mathrm{PP}}\right]_{2,2},...,\left[\mathbf{Z}_{\mathrm{PP}}\right]_{Q,Q}\right)$
and define a diagonal matrix as $\mathbf{Y}_{\mathrm{D}}\left(\mathbf{b}\right)=\left(\mathbf{Z}_{\mathrm{PP,D}}+\mathbf{Z}_{\mathrm{L}}\left(\mathbf{b}\right)\right)^{-1}$.
From \eqref{eq:Pattern}, we focus on the radiation pattern of pixel
antenna with a unit feeding current at the antenna port, i.e. $i_{A}=1$,
which can be written as
\begin{align}
\tilde{\mathbf{e}}\left(\mathbf{b}\right) & =\mathbf{e}_{\mathrm{A}}-\mathbf{E}_{\mathrm{P}}\left(\mathbf{Z}_{\mathrm{PP}}+\mathbf{Z}_{\mathrm{L}}\left(\mathbf{b}\right)\right)^{-1}\mathbf{z}_{\mathrm{PA}}\nonumber \\
 & =\mathbf{e}_{\mathrm{A}}-\mathbf{E}_{\mathrm{P}}\left(\mathbf{Z}_{\mathrm{PP}}-\mathbf{Z}_{\mathrm{PP,D}}+\mathbf{Y}_{\mathrm{D}}^{-1}\left(\mathbf{b}\right)\right)^{-1}\mathbf{z}_{\mathrm{PA}}.\label{eq:Diagonalization}
\end{align}
In most cases, there is at least one switch that is in the off state
(i.e. the load impedance is open-circuited) so that at least one diagonal
entry in $\mathbf{Z}_{\mathrm{L}}\left(\mathbf{b}\right)$ is extremely
large impedance $z_{\mathrm{oc}}$. Therefore, we can make the approximation
that $\left\Vert \mathbf{Y}_{\mathrm{D}}^{-1}\left(\mathbf{b}\right)\right\Vert _{F}=\left\Vert \mathbf{Z}_{\mathrm{PP,D}}+\mathbf{Z}_{\mathrm{L}}\left(\mathbf{b}\right)\right\Vert _{F}\thickapprox\left\Vert \mathbf{Z}_{\mathrm{L}}\left(\mathbf{b}\right)\right\Vert _{F}\gg\left\Vert \mathbf{Z}_{\mathrm{PP}}-\mathbf{Z}_{\mathrm{PP,D}}\right\Vert _{F}$
which means that $\left\Vert \mathbf{Y}_{\mathrm{D}}^{-1}\left(\mathbf{b}\right)\right\Vert _{F}$
is extremely large. Defining $\mathbf{Z}_{\mathrm{PP,O}}=\mathbf{Z}_{\mathrm{PP}}-\mathbf{Z}_{\mathrm{PP,D}}$
as the matrix consisting of all the off-diagonal entries in $\mathbf{Z}_{\mathrm{PP}}$
(i.e. the mutual impedance), we can use perturbation theory to expand
the matrix inversion term in \eqref{eq:Diagonalization}, written
as
\begin{equation}
\left(\mathbf{Z}_{\mathrm{PP,O}}+\mathbf{Y}_{\mathrm{D}}^{-1}\left(\mathbf{b}\right)\right)^{-1}=\mathbf{Y}_{\mathrm{D}}\left(\mathbf{b}\right)\sum_{i=0}^{\infty}\left(-\mathbf{Z}_{\mathrm{PP,O}}\mathbf{Y}_{\mathrm{D}}\left(\mathbf{b}\right)\right)^{i}.\label{eq:Diagonalization 1}
\end{equation}
We can further notice that the higher order terms with $i\geqslant1$
are much smaller than the term with $i=0$ due to an extremely small
$\left\Vert \mathbf{Y}_{\mathrm{D}}\left(\mathbf{b}\right)\right\Vert _{F}$.
Accordingly, we can take the first term in \eqref{eq:Diagonalization 1}
into \eqref{eq:Diagonalization} as
\begin{equation}
\tilde{\mathbf{e}}\left(\mathbf{b}\right)=\mathbf{e}_{\mathrm{A}}-\mathbf{E}_{\mathrm{P}}\mathbf{Y}_{\mathrm{D}}\left(\mathbf{b}\right)\mathbf{z}_{\mathrm{PA}}.\label{eq:Diagonalization 2}
\end{equation}
Defining $\mathbf{y}_{\mathrm{D}}\left(\mathbf{b}\right)=\left[[\mathbf{Y}_{\mathrm{D}}\left(\mathbf{b}\right)]_{1,1},...,[\mathbf{Y}_{\mathrm{D}}\left(\mathbf{b}\right)]_{Q,Q}\right]^{T}$
and $\mathbf{Z}_{\mathrm{PA}}=\mathrm{diag}\left(\mathbf{z}_{\mathrm{PA}}\right)$,
we can rewrite \eqref{eq:Diagonalization 2} as
\begin{align}
\tilde{\mathbf{e}}\left(\mathbf{b}\right)= & \mathbf{e}_{\mathrm{A}}-\mathbf{E}_{\mathrm{P}}\mathbf{Z}_{\mathrm{PA}}\mathbf{y}_{\mathrm{D}}\left(\mathbf{b}\right),\label{eq:Diagonalization 3}
\end{align}
It can be observed from \eqref{eq:Diagonalization 3} that the mutual
impedance between the single antenna port and the $Q$ pixel ports
$\mathbf{Z}_{\mathrm{PA}}$ is a weighting matrix imposed on the open-circuit
radiation patterns of pixel ports $\mathbf{E}_{\mathrm{P}}$. $\mathbf{y}_{\mathrm{D}}\left(\mathbf{b}\right)$
is utilized to activate the selected pixel ports. That is, $\left[\mathbf{Y}_{\mathrm{D}}\left(\mathbf{b}\right)\right]_{q,q}$
is zero when the $q$th RF switch is off (i.e. the $q$th pixel port
is open-circuited) so that the $q$th open-circuit radiation pattern
in $\mathbf{E}_{\mathrm{P}}$ is not radiating. While $\left[\mathbf{Y}_{\mathrm{D}}\left(\mathbf{b}\right)\right]_{q,q}=\left[\mathbf{Z}_{\mathrm{PP}}\right]_{q,q}^{-1}$
is the inverse of the self impedance of the $q$th pixel port when
the $q$th RF switch is on (i.e. the $q$th pixel port is short-circuited).
Thus we can rewrite \eqref{eq:Diagonalization 3} as 
\begin{equation}
\tilde{\mathbf{e}}\left(\mathbf{b}\right)=\mathbf{e}_{\mathrm{A}}-\mathbf{E}_{\mathrm{P}}\mathbf{Z}_{\mathrm{PA}}\mathbf{Z}_{\mathrm{PP,D}}^{-1}\left(\mathbf{u}-\mathbf{b}\right),\label{eq:Diagonalization 4}
\end{equation}
where $\mathbf{u}=\left[1,1,...,1\right]^{T}\in\mathbb{N}^{Q\times1}$
is a vector with all entries being unity.

We can observe from \eqref{eq:Diagonalization 4} that the radiation
pattern of the pixel antenna can be approximately obtained by linearly
combining the weighted open-circuit radiation patterns of pixel ports
$\mathbf{E}_{\mathrm{P}}\mathbf{Z}_{\mathrm{PA}}\mathbf{Z}_{\mathrm{PP,D}}^{-1}$
with the open-circuit radiation patterns of a single antenna port
$\mathbf{e}_{\mathrm{A}}$. From the circuit perspective, the term
$\mathbf{Z}_{\mathrm{PA}}\mathbf{Z}_{\mathrm{PP,D}}^{-1}$ can be
viewed as the mutual coupling strength between the antenna port and
pixel ports. Therefore, by adjusting the antenna coder $\mathbf{b}$,
the RF switch state is either on and off, so that the corresponding
open-circuit radiation pattern will be either activated or de-activated
in constructing the radiation pattern of pixel antenna, which shows
that the radiation pattern of the pixel antenna can be approximately
a binary linear combinations of all open-circuit radiation patterns
coded by the antenna coder.

\subsection{Analysis for Effective Aerial Degrees-of-Freedom}

Theoretically, the EADoF $R=\mathrm{rank}\left(\mathbf{E}_{\mathrm{oc}}\right)$
refers to the number of orthonormal basis radiation patterns that
a pixel antenna can provide, which is an upper bound on the codebook
size when modulating the pattern coder $\mathbf{w}\left(\mathbf{b}_{\mathrm{C},p}\right)$
in \eqref{eq:pattern coder SSK} to exploit the spatial multiplexing.
However, in practice when we numerically perform the singular value
decomposition for $\mathbf{E}_{\mathrm{oc}}$, there generally exists
some very small singular values, which poses a challenge to accurately
evaluate the EADoF.

To overcome this challenge, we analyze the EADoF from the perspective
of radiated power. It is known that each singular value in $\mathbf{\Sigma}$
is associated to one basis radiation pattern in the beamspace domain
and the basis radiation patterns associated with those very small
singular values cannot be effectively excited. That is, $\sigma_{i}^{2}$
implies the capability to radiate power for the $i$th basis radiation
pattern. Therefore, we define the cumulative distribution function
as
\begin{equation}
F_{i}=\frac{\sum_{j=1}^{i}\sigma_{j}^{2}}{\sum_{j=1}^{Q+1}\sigma_{j}^{2}},\:i=1,2,...,Q+1,\label{eq:cumulative power}
\end{equation}
which satisfies $0\leqslant F_{i}\leqslant1$ and characterizes the
sum power of the first $i$ basis patterns in the radiation of the
pixel antenna. By setting the threshold as $T$, we can find the EADoF
as
\begin{equation}
R=\underset{F_{i}\geqslant T}{\mathrm{argmin}}\:i,\label{eq:pattern basis}
\end{equation}
which means that the chosen basis radiation patterns contribute most
of the radiated power for the pixel antenna.

\subsection{Antenna Coding Optimization}

In this subsection, we propose an efficient optimization method for
designing the antenna coders for exploiting the spatial multiplexing.
With the analysis above, we aim to optimize $P$ $\left(P\leqslant R\right)$
antenna coders $\mathbf{b}_{\mathrm{C},p}$ for $p=1,2,...,P$ to
form the codebook $\mathcal{B}=\left\{ \mathbf{b}_{\mathrm{C},1},\mathbf{b}_{\mathrm{C},2},...,\mathbf{b}_{\mathrm{C},P}\right\} $
so that the pattern coder $\mathbf{W}\left(\mathbf{B}\right)$ can
be modulated by selecting among the $P$ basis radiation patterns
of the pixel antenna for enhancing the spectral efficiency.

We can observe that the pattern coders of two basis radiation patterns
satisfy $\mathbf{w}^{H}\left(\mathbf{b}_{\mathrm{C},j}\right)\mathbf{w}\left(\mathbf{b}_{\mathrm{C},k}\right)=\delta_{j,k}$,
for $j,k=1,2,...,P$, where $\delta_{j,k}$ is the Kronecker delta
function (0 if $j\neq k$ and 1 if $j=k$). Therefore, the correlation
coefficient between two pattern coders can be utilized to evaluate
their similarity, which is defined as
\begin{align}
\rho_{j,k}\left(\mathbf{b}_{\mathrm{C},j},\mathbf{b}_{\mathrm{C},k}\right)= & \mathbf{w}^{H}\left(\mathbf{b}_{\mathrm{C},j}\right)\mathbf{w}\left(\mathbf{b}_{\mathrm{C},k}\right),\label{eq:CorrCoef}
\end{align}
where $\rho_{j,k}$ satisfies $0\leq\left|\rho_{j,k}\right|\leq1$.
When $\left|\rho_{jk}\right|=0$, $\mathbf{w}\left(\mathbf{b}_{\mathrm{C},j}\right)$
and $\mathbf{w}\left(\mathbf{b}_{\mathrm{C},k}\right)$ are orthogonal
to each other and can be associated to different basis radiation patterns.
Leveraging the correlation coefficient \eqref{eq:CorrCoef}, the optimization
problem can be formulated as
\begin{align}
\underset{\mathbf{b}_{\mathrm{C},p},\forall p}{\mathrm{min}}\:\: & \frac{2}{P\left(P-1\right)}\stackrel[k=j+1]{P}{\sum}\stackrel[j=1]{P}{\sum}\left|\rho_{j,k}\left(\mathbf{b}_{\mathrm{C},j},\mathbf{b}_{\mathrm{C},k}\right)\right|^{2}\label{eq:problem}\\
\mathrm{s.t.}\:\:\:\:\: & \mathbf{b}_{\mathrm{C},p}\in\left\{ 0,1\right\} ^{Q},\forall p,\label{eq:subj}
\end{align}
where the objective \eqref{eq:problem} calculates the mean correlation
between all pairs of pattern coders. To optimize the binary variables
for pixel antennas, multiple optimization methods including successive
exhaustive Boolean optimization \cite{7762757}, perturbation sensitivity
\cite{9491941} and adjoint method \cite{10035928} have been proposed.
In this work, we use the genetic algorithm (GA) \cite{Zhang2022},
\cite{8862255} to solve the binary optimization problem \eqref{eq:problem}-\eqref{eq:subj},
so as to obtain the optimal codebook denoted as $\mathcal{B}^{\star}=\left\{ \mathbf{b}_{\mathrm{C},1}^{\star},\mathbf{b}_{\mathrm{C},2}^{\star},...,\mathbf{b}_{\mathrm{C},P}^{\star}\right\} $.
The optimal codebook $\mathcal{B}^{\star}$ achieves the minimum mean
correlation, denoted as $g^{\star}$, which however requires $Q$
RF switches to implement the pixel antenna. In the following subsection,
the optimal codebook $\mathcal{B}^{\star}$ with the minimum mean
correlation $g^{\star}$ will be used as a baseline for minimizing
the number of RF switches.

\subsection{Minimizing Number of RF Switches}

In practical implementations, we wish to minimize the number of RF
switches to reduce the circuit complexity of the pixel antenna, while
maintaining orthogonality among the $P$ pattern coders. To that end,
if the switch state at the $q$th pixel port for all the antenna coders
$\mathbf{b}_{\mathrm{C},p},\forall p$ are the same, that is $\left[\mathbf{b}_{\mathrm{C},1}\right]_{q}=\ldots=\left[\mathbf{b}_{\mathrm{C},P}\right]_{q}$,
then we can replace the $q$th switch with a fixed short circuit (through
hardwire) or open circuit. Therefore, we only need to use switches
across the pixel ports which are different for the antenna coders
$\mathbf{b}_{\mathrm{C},p},\forall p$. In addition, energy efficiency
can also be improved since power consumption by the switches is reduced.
However, it is NP-hard to solve the problem \eqref{eq:problem}-\eqref{eq:subj}
while minimizing the number of RF switches, because the switch states
of all antenna coders, the pixel ports with RF switches, and the pixel
ports with fixed open/short circuit need to be optimized jointly.

\begin{algorithm}[t]
\caption{Minimizing number of RF switches.\label{alg:AO}}

\textbf{Input:} $\mathcal{B}^{\star}$, $g^{\star}$, $\Delta g$,
$\mathcal{Q}$;

$\quad${\scriptsize 1:} \textbf{Initialization:} $\mathcal{B}^{\left(0\right)}$,
$\mathcal{V}^{\text{\ensuremath{\left(0\right)}}}$, $\mathcal{W}^{\left(0\right)}$,
$i=1$;

$\quad${\scriptsize 2:} \textbf{repeat}

$\quad${\scriptsize 3:} $\:\:\:\:$\textbf{if} $\exists v\in\mathcal{V}^{\left(i-1\right)}$,
$\left[\mathbf{b}_{\mathrm{C},1}^{\left(i-1\right)}\right]_{v}=\cdots=\left[\mathbf{b}_{\mathrm{C},P}^{\left(i-1\right)}\right]_{v}$

$\quad${\scriptsize 4:} $\:\:\:\:\:\:\:\:$\textbf{for} $\forall v\in\mathcal{V}^{\left(i-1\right)}$
that $\left[\mathbf{b}_{\mathrm{C},1}^{\left(i-1\right)}\right]_{v}=\ldots=\left[\mathbf{b}_{\mathrm{C},P}^{\left(i-1\right)}\right]_{v}$

$\quad${\scriptsize 5:} $\:\:\:\:$$\:\:\:\:$$\:\:\:\:$Obtain $\mathcal{W}^{\left(i\right)}=\left\{ \mathcal{W}^{\left(i-1\right)},v\right\} $;

$\quad${\scriptsize 6:} $\:\:\:\:\:\:\:\:$\textbf{end}

$\quad${\scriptsize 7:} $\:\:\:\:\:\:\:\:$Obtain $\mathcal{V}^{\left(i\right)}=\mathcal{Q}\setminus\mathcal{W}^{\left(i\right)}$;

$\quad${\scriptsize 8:} $\:\:\:\:\:\:\:\:$Obtain $\mathcal{B}^{\left(i\right)}=\mathcal{B}^{\left(i-1\right)}$;

$\quad${\scriptsize 9:} $\:\:\:\:$\textbf{else}

$\,\,\,\,${\scriptsize 10:} $\:\:\:\:$$\:\:\:\:$Find $\bar{v}^{\left(i\right)}$
and $b^{\left(i\right)}$ by \eqref{eq:problem-2-1}-\eqref{eq: b=00003D0,1};

$\,\,\,\,${\scriptsize 11:} $\:\:\:\:$$\:\:\:\:$Obtain $\mathcal{W}^{\left(i\right)}=\left\{ \mathcal{W}^{\left(i-1\right)},\bar{v}^{\left(i\right)}\right\} $,
$\mathcal{V}^{\left(i\right)}=\mathcal{Q}\setminus\mathcal{W}^{\left(i\right)}$;

$\,\,\,\,${\scriptsize 12:} $\:\:\:\:$$\:\:\:\:$Obtain $\left[\mathbf{b}_{\mathrm{C},p}^{\left(i\right)}\right]_{\bar{v}^{\left(i\right)}}=b^{\left(i\right)}$,
$\forall p$;

$\,\,\,\,${\scriptsize 13:} $\:\:\:\:$$\:\:\:\:$Obtain $\left[\mathbf{b}_{\mathrm{C},p}^{\left(i\right)}\right]_{q}=\left[\mathbf{b}_{\mathrm{C},p}^{\left(i-1\right)}\right]_{q}$,
$\forall q\in\mathcal{Q}\setminus\bar{v}^{\left(i\right)}$, $\forall p$;

$\,\,\,\,${\scriptsize 14:} $\:\:\:\:$\textbf{end}

$\,\,\,\,${\scriptsize 15:} $\:\:\:\:$Find $\left[\mathbf{b}_{\mathrm{C},p}^{\left(i\right)}\right]_{v}$,
$\forall v\in\mathcal{V}^{\left(i\right)}$, $\forall p$, and $g^{\left(i\right)}$
by \eqref{eq:problem-1}-\eqref{eq: sub-remain fixed};

$\,\,\,\,${\scriptsize 16:} $\:\:\:\:$Update $\mathcal{B}^{\left(i\right)}$
by $\left[\mathbf{b}_{\mathrm{C},p}^{\left(i\right)}\right]_{v},\forall v\in\mathcal{V}^{\left(i\right)},\forall p$
;

$\quad$$\quad$$\:\:\:\:$$i=i+1$;

$\,\,\,\,${\scriptsize 17:} \textbf{until} $g^{\left(i\right)}-g^{\star}\geqslant\Delta g$;

$\,\,\,\,${\scriptsize 18:} Obtain $N_{v}=\mathrm{card}\left(\mathcal{V}^{\left(i-1\right)}\right)$;

\textbf{Output:} $\mathcal{B}^{\left(i-1\right)},\mathcal{V}^{\left(i-1\right)},\mathcal{W}^{\left(i-1\right)}$,
and $N_{v}$;
\end{algorithm}
To minimize the number of RF switches, we propose an iterative algorithm
to alternatively reduce the pixel ports with RF switches and optimize
the antenna coder. The optimal codebook obtained by GA as proposed
in Section IV.C is chosen as the starting point of the iterative algorithm
$\mathcal{B}^{\left(0\right)}=\mathcal{B}^{\star}$, that is to say
we aim to minimize the RF switches based on the optimal codebook $\mathcal{B}^{\star}$.
We use a set $\mathcal{Q}=\left\{ 1,2,...,Q\right\} $ to collect
all indices of pixel ports. We also define $\mathcal{V}^{\left(0\right)}=\left\{ 1,2,...,Q\right\} $
and $\mathcal{W}^{\left(0\right)}=\mathcal{Q}\setminus\mathcal{V}^{\left(0\right)}$
as the sets of indices of pixel ports with RF switches and with fixed
open/short circuit at the starting point of the iterative algorithm.

At the $i$th iteration, we first update $\mathcal{V}^{\left(i\right)}$
and $\mathcal{W}^{\left(i\right)}$ based on $\mathcal{B}^{\left(i-1\right)}$
to reduce the number of switches. If there exists the same switch
state for all the antenna coders in $\mathcal{B}^{\left(i-1\right)}$
at a certain pixel port indexed in $\mathcal{V}^{\left(i-1\right)}$,
i.e. $\exists v\in\mathcal{V}^{\left(i-1\right)}$, $\left[\mathbf{b}_{\mathrm{C},1}^{\left(i-1\right)}\right]_{v}=\ldots=\left[\mathbf{b}_{\mathrm{C},P}^{\left(i-1\right)}\right]_{v}$,
we update the set of indices of pixel ports with fixed open/short
circuit and with switches by $\mathcal{W}^{\left(i\right)}=\left\{ \mathcal{W}^{\left(i-1\right)},v\right\} $
and $\mathcal{V}^{\left(i\right)}=\mathcal{Q}\setminus\mathcal{W}^{\left(i\right)}$,
respectively, while the codebook remains the same, that is $\mathcal{B}^{\left(i\right)}\triangleq\left\{ \mathbf{b}_{\mathrm{C},1}^{\left(i\right)},\ldots,\mathbf{b}_{\mathrm{C},P}^{\left(i\right)}\right\} =\mathcal{B}^{\left(i-1\right)}$.
On the other hand, if there does not exist the same switch state for
all the antenna coders, we select one index in $\mathcal{V}^{\left(i-1\right)}$
and force the corresponding pixel port to have the same switch state,
so that the switch can be replaced by the fixed open/short circuit.
Regarding which pixel port to be selected, this is conducted by sequentially
searching each of the pixel ports in $\mathcal{V}^{\left(i-1\right)}$
with the same switch state, either on or off, to minimize the mean
correlation, which can be formulated as
\begin{align}
\underset{\bar{v},b}{\mathrm{min}}\:\:\, & \frac{2}{P\left(P-1\right)}\stackrel[k=j+1]{P}{\sum}\stackrel[j=1]{P}{\sum}\left|\rho_{j,k}\left(\mathbf{b}_{\mathrm{C},j},\mathbf{b}_{\mathrm{C},k}\right)\right|^{2}\label{eq:problem-2-1}\\
\mathrm{s.t.}\:\:\: & \left[\mathbf{b}_{\mathrm{C},p}\right]_{v}=\left[\mathbf{b}_{\mathrm{C},p}^{\left(i-1\right)}\right]_{v},\forall v\in\left(\mathcal{V}^{\left(i-1\right)}\setminus\bar{v}\right),\forall p,\label{eq:subj-2-1}\\
 & \left[\mathbf{b}_{\mathrm{C},p}\right]_{w}=\left[\mathbf{b}_{\mathrm{C},p}^{\left(i-1\right)}\right]_{w},\forall w\in\mathcal{W}^{\left(i-1\right)},\forall p,\\
 & \left[\mathbf{b}_{\mathrm{C},p}\right]_{\bar{v}}=b,\forall p,\\
 & \bar{v}\in\mathcal{V}^{\left(i-1\right)},\\
 & b\in\left\{ 0,1\right\} ,\label{eq: b=00003D0,1}
\end{align}
which can be solved by sequentially searching $\bar{v}$ in the set
$\mathcal{V}^{\left(i-1\right)}$ with the binary variable $b$. The
optimal index and switch state from problem \eqref{eq:problem-2-1}
to \eqref{eq: b=00003D0,1} is denoted as $\bar{v}^{\left(i\right)}$
and $b^{\left(i\right)}$. Accordingly, the set of indices of pixel
ports with fixed open/short circuit and with RF switches are updated
by $\mathcal{W}^{\left(i\right)}=\left\{ \mathcal{W}^{\left(i-1\right)},\bar{v}^{\left(i\right)}\right\} $
and $\mathcal{V}^{\left(i\right)}=\mathcal{Q}\setminus\mathcal{W}^{\left(i\right)}$,
respectively, while the codebook $\mathcal{B}^{\left(i\right)}$ is
obtained by setting $\left[\mathbf{b}_{\mathrm{C},p}^{\left(i\right)}\right]_{\bar{v}^{\left(i\right)}}=b^{\left(i\right)}$
and $\left[\mathbf{b}_{\mathrm{C},p}^{\left(i\right)}\right]_{q}=\left[\mathbf{b}_{\mathrm{C},p}^{\left(i-1\right)}\right]_{q}$,
$\forall q\in\mathcal{Q}\setminus\bar{v}^{\left(i\right)}$, $\forall p$.

After updating $\mathcal{V}^{\left(i\right)}$ and $\mathcal{W}^{\left(i\right)}$,
we next wish to optimize the switch states at the pixel ports within
$\mathcal{V}^{\left(i\right)}$ for the antenna coders to minimize
the mean correlation, formulated as
\begin{align}
\underset{\mathbf{b}_{\mathrm{C},p},\forall p}{\mathrm{min}}\:\: & \frac{2}{P\left(P-1\right)}\stackrel[k=j+1]{P}{\sum}\stackrel[j=1]{P}{\sum}\left|\rho_{j,k}\left(\mathbf{b}_{\mathrm{C},j},\mathbf{b}_{\mathrm{C},k}\right)\right|^{2}\label{eq:problem-1}\\
\mathrm{s.t.}\:\:\:\:\: & \left[\mathbf{b}_{\mathrm{C},p}\right]_{v}\in\left\{ 0,1\right\} ,\forall v\in\mathcal{V}^{\left(i\right)},\forall p,\label{eq:subj-1-1}\\
 & \left[\mathbf{b}_{\mathrm{C},p}\right]_{w}=\left[\mathbf{b}_{\mathrm{C},p}^{\left(i\right)}\right]_{w},\forall w\in\mathcal{W}^{\left(i\right)},\forall p,\label{eq: sub-remain fixed}
\end{align}
where the entries of antenna coder $\left[\mathbf{b}_{\mathrm{C},p}\right]_{w}$,
$\forall w\in\mathcal{W}^{\left(i\right)}$, $\forall p$, remain
fixed. Using $\mathcal{B}^{\left(i\right)}$ obtained in the first
step of the $i$th iteration as one of the initial population, we
can use GA again to solve the problem \eqref{eq:problem-1}-\eqref{eq: sub-remain fixed}.
The switch states for the pixel ports within $\mathcal{V}^{\left(i\right)}$
optimized by GA is denoted as $\left[\mathbf{b}_{\mathrm{C},p}^{\left(i\right)}\right]_{v}$,
$\forall v\in\mathcal{V}^{\left(i\right)}$, $\forall p$, and accordingly
the codebook at the $i$th iteration $\mathcal{B}^{\left(i\right)}$
is also updated where the optimized mean correlation value is $g^{\left(i\right)}$.

By iteratively updating the set of indices of pixel ports with RF
switches and optimizing the antenna coder, the number of RF switches
can be gradually reduced. Nevertheless, it should be noted that reducing
an RF switch by forcing the antenna coder at a pixel port to be the
same, as part of the first step in the $i$th iteration, is achieved
at the expense of a potential increase of the mean correlation. Therefore,
the mean correlation $g^{\left(i\right)}$ can increase with the iteration
$i$. Because our aim is to minimize the number of RF switches while
maintaining the mean correlation lower than an acceptable value to
achieve the approximate orthogonality among the $P$ pattern coders,
we set the stopping criterion of the iterative algorithm by
\begin{equation}
g^{\left(i\right)}-g^{\star}\geqslant\Delta g,\label{eq:Stopping Criterion}
\end{equation}
where $\Delta g$ is the stopping threshold denoting as the difference
between the mean correlation of the $i$th iteration $g^{\left(i\right)}$
and the baseline mean correlation $g^{\star}$ which requires the
full $Q$ RF switches. When \eqref{eq:Stopping Criterion} is not
satisfied, it means that we have successfully reduced the number of
switches at the $i$th iteration while maintaining the mean correlation
lower than an acceptable value. When \eqref{eq:Stopping Criterion}
is satisfied, the mean correlation $g^{\left(i\right)}$ is too large,
meaning that we cannot achieve the orthogonality among the $P$ pattern
coders, even approximately, by $\mathcal{B}^{\left(i\right)}$ optimized
at the $i$th iteration. Thus, the iterative algorithm stops and the
results optimized in the $(i-1)$th iteration, i.e. $\mathcal{B}^{\left(i-1\right)}$,
$\mathcal{V}^{\left(i-1\right)}$, and $\mathcal{W}^{\left(i-1\right)}$,
will be used as the final codebook design with the minimized number
of RF switches given by $N_{v}=\mathrm{card}\left(\mathcal{V}^{\left(i-1\right)}\right)$.

Algorithm \ref{alg:AO} summarizes the overall iterative algorithm
for designing the antenna coder with minimizing number of RF switches.
The performance of the algorithm will be evaluated in the next section.

\section{Numerical Simulation }

\begin{figure}[t]
\begin{centering}
\includegraphics[width=8cm]{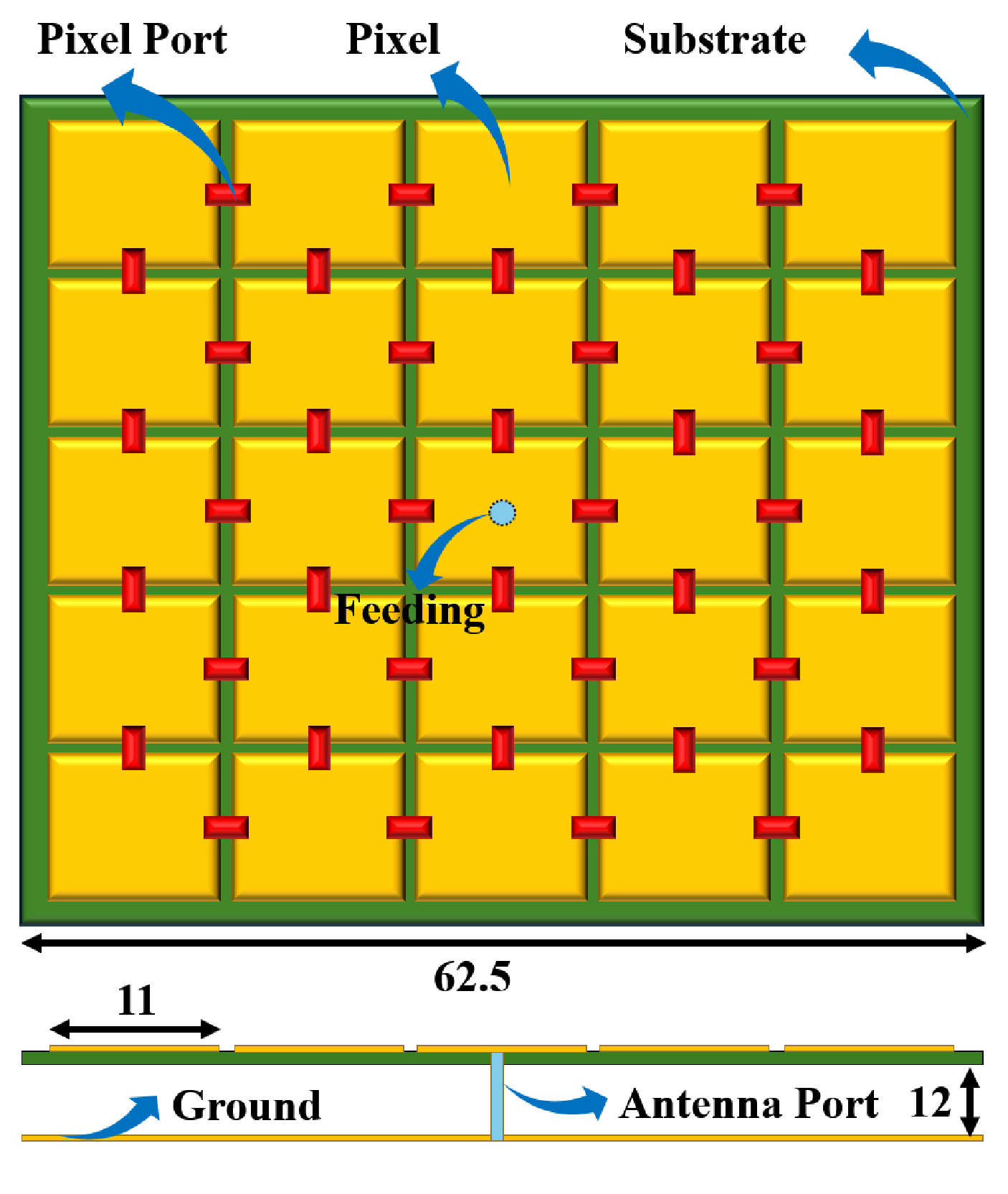}
\par\end{centering}
\caption{Plan and elevation views of the pixel antenna design (Geometry and
dimension unit: mm). \label{fig:Design}}
\end{figure}
In this section, we firstly provide a design example for the pixel
antenna and analyze its DoF in the beamspace domain. Then we use the
proposed optimization algorithms in Section IV to obtain the antenna
coders to implement the orthonormal radiation patterns with the minimized
number of RF switches. Using the optimized radiation patterns of the
pixel antenna, we evaluate the performance of the MIMO systems enhanced
by antenna coding.

\subsection{Pixel Antenna Design}

We design an exemplary pixel antenna working at 2.4 GHz where the
wavelength is 125 mm. The structure and geometry of the pixel antenna
design are shown in Fig. \ref{fig:Design} where a $5\times5$ pixel
array is mounted on a substrate with thickness of 1.524 mm. The pixels
are made of copper which has electric conductivity of $5.8\times10^{7}$
S/m and the substrate is made of Rogers 4003C which has loss tangent
of 0.0027 and permittivity of 3.55. The side length of pixel and substrate
are 11 mm and 62.5 mm. There are $Q=40$ pixel ports placed across
each pair of adjacent pixels, which can be implemented by RF switches
or fixed short circuit (through hardwire) or open circuit. To implement
the antenna port, a copper plane is placed underneath the pixel array
with 12 mm air gap to the substrate so that a feeding probe can be
used to connect the ground and the center pixel. It should be noted
that this pixel antenna design can be regarded as discretizing a conventional
square patch antenna into a pixel array. We use a full electromagnetic
(EM) solver, CST studio suite \cite{CST}, to simulate this 41-port
pixel antenna to obtain the impedance matrix $\mathbf{Z}$ and open-circuit
radiation pattern matrix $\mathbf{E}_{\mathrm{oc}}$. It should be
noted that the full EM simulation only needs to be performed once
for the pixel antenna because any pattern coder excited by any antenna
coder and corresponding radiation pattern can then be found by using
\eqref{eq:Pattern} and \eqref{eq:pattern coder}. Thus, the computational
complexity can be reduced enormously as full EM simulation is not
needed during the antenna coding optimization and performance analysis.

\begin{figure}[t]
\begin{centering}
\includegraphics[width=8.5cm]{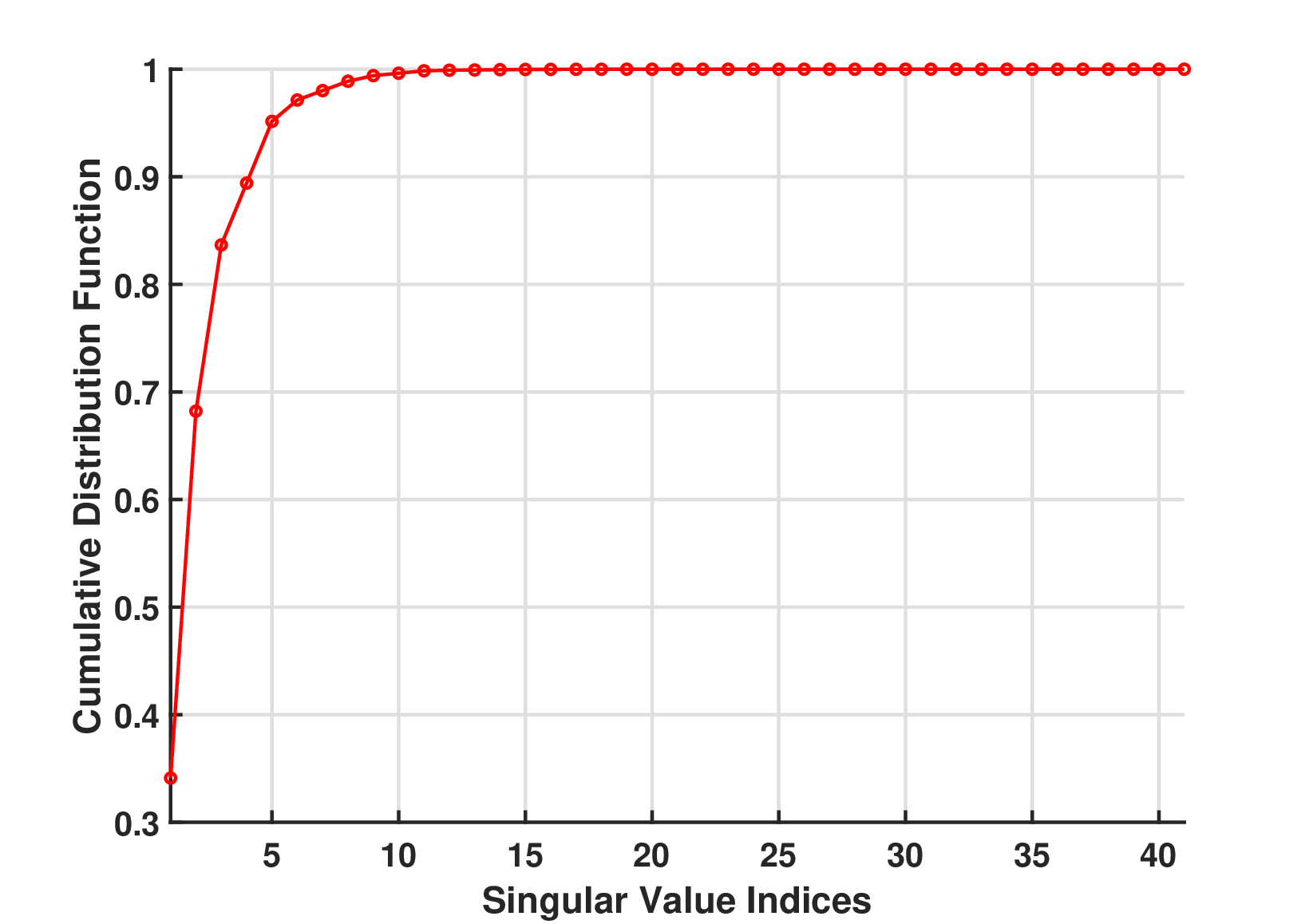}
\par\end{centering}
\caption{Cumulative distribution function for the proposed pixel antenna. \label{fig:EADoF}}
\end{figure}
In addition, we consider the beamspace with 3D uniform power angular
spectrum. The angular resolution is $5^{\circ}$ so that we have $K=2664$
sampling points. We also define the SNR as $\mathrm{SNR}=\frac{\text{\ensuremath{P_{T}}}}{P_{L}N_{0}}$
where $P_{T}$ denotes the transmit power, $P_{L}$ denotes the path
loss, and $N_{0}$ is the noise power. In the simulation, we assume
$P_{L}=100$ dB and $N_{0}=-80$ dBm.

We analyze the EADoF that can be provided by the proposed pixel antenna
in the beamspace domain. The cumulative distribution function for
the pixel antenna, as calculated in \eqref{eq:cumulative power},
is shown in Fig. \ref{fig:EADoF}. It can be straightforwardly observed
that the radiated power of most basis radiation patterns are close
to zero so that only few basis can be utilized for antenna coding
design. In this work, by setting the threshold as $0.995$, we can
find the EADoF from \eqref{eq:pattern basis} is $R=8$. That is,
99.5\% power of the radiation patterns of the pixel antenna are contributed
by the first $R=8$ basis. Therefore, we can conclude that the EADoF
provided by the pixel antenna is 8. In the following optimization
and analysis, we use such EADoF for guiding the antenna coding design.

\subsection{Pixel Antenna Optimization}

\begin{figure}[t]
\begin{centering}
\includegraphics[width=8.5cm]{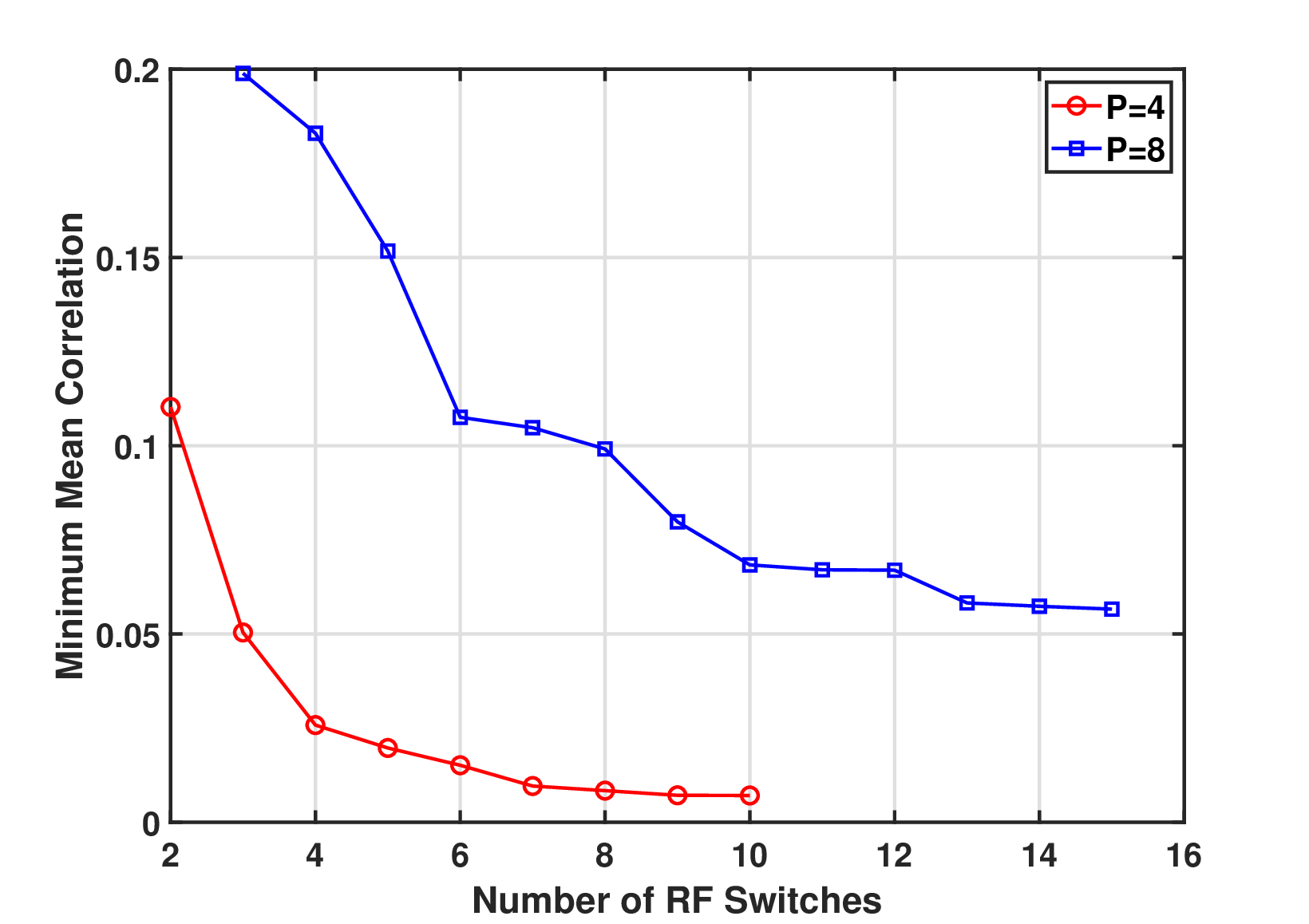}
\par\end{centering}
\caption{Minimum mean correlation versus the number of RF switches.\label{fig:Obj}}
\end{figure}

In this subsection, we present the codebook design for antenna coding
which implements the $P$ orthonormal basis radiation patterns to
modulate the pattern coders for spectral efficiency enhancement. Specifically,
we aim to optimize codebooks with $P=4$ and 8 antenna coders to achieve
4 and 8 basis radiation patterns. We first use GA to solve the problem
\eqref{eq:problem}-\eqref{eq:subj} to find the baseline value of
the mean correlation $g^{\star}$, which is 0.005 and 0.05 for $P=4$
and 8 cases, respectively. Next, using Algorithm 1 proposed in Section
IV.D, we can find the minimum mean correlation for different numbers
of RF switches as shown in Fig. \ref{fig:Obj}. We can find that the
minimum mean correlation increases when we decrease the number of
RF switches, which is because the reduced RF switches limit the reconfigurability
of pixel antennas. Thus, there is a tradeoff between the orthogonality
among pattern coders and the number of RF switches. By setting the
stopping threshold as $\Delta g=$ $0.05$ and $0.1$ for $P=4$ and
8 cases, the least number of RF switches to satisfy the mean correlation
threshold \eqref{eq:Stopping Criterion} is 3 and 5, respectively.
In $P=4$ case, compared with the 2 RF switches (the least number
of RF switches $\textrm{log}_{2}P$ to achieve 4 different patterns)
which achieves the minimum mean correlation of around 0.11, the 3
RF switches achieves lower mean correlation of around 0.05 with requiring
only one extra RF switch. Besides, in $P=8$ case, compared with the
$\textrm{log}_{2}P=3$ RF switches with the minimum mean correlation
of around 0.2, the 5 RF switches achieves lower mean correlation of
around 0.15 with requiring only two extra RF switches. Therefore,
these results show that the proposed Algorithm 1 can effectively minimize
the number of RF switches while maintaining an approximate orthogonality
between pattern coders.

\begin{figure}[t]
\begin{centering}
\includegraphics[width=7cm]{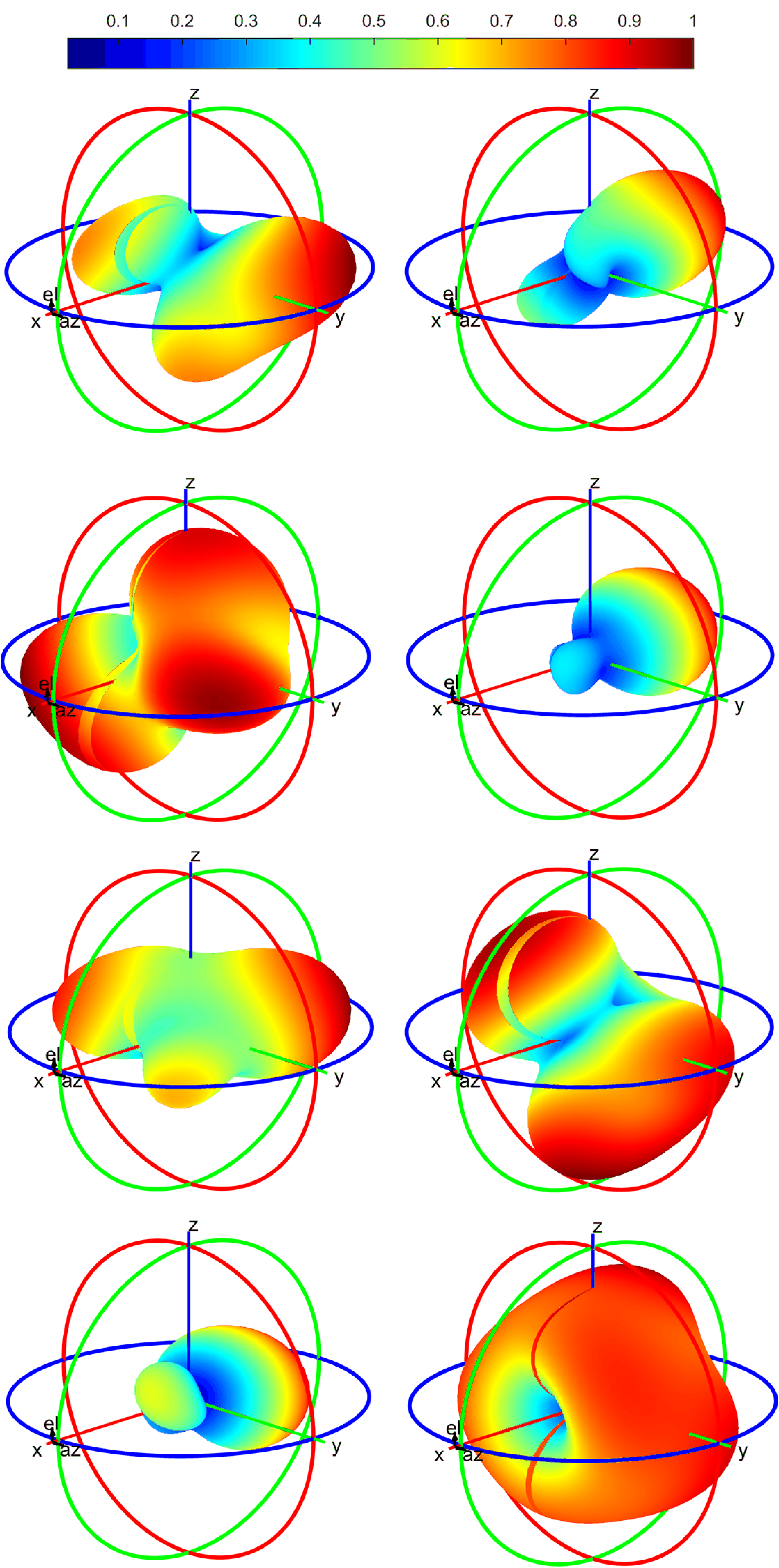}
\par\end{centering}
\caption{Radiation patterns of the proposed pixel antenna with the optimized
codebook using 5 RF switches in $P=8$ case. \label{fig:Pattern}}
\end{figure}
In Fig. \ref{fig:Pattern}, we provide the radiation patterns of the
proposed pixel antenna with the optimized codebook using 5 RF switches
in $P=8$ case for reference. These are plotted in the full 3D sphere
in the far field. We can notice that the radiation patterns are quite
different due to the optimized different antenna coders. It should
be noted that these radiation patterns are close to orthogonal so
that selecting among these radiation patterns can transmit additional
information bits in the beamspace domain. Radiation patterns for $P=4$
case and the other numbers of RF switches follow a similar trend so
we omit them.

\subsection{Performance of MIMO System with Antenna Coding}

\begin{figure}[t]
\begin{centering}
\includegraphics[width=8.5cm]{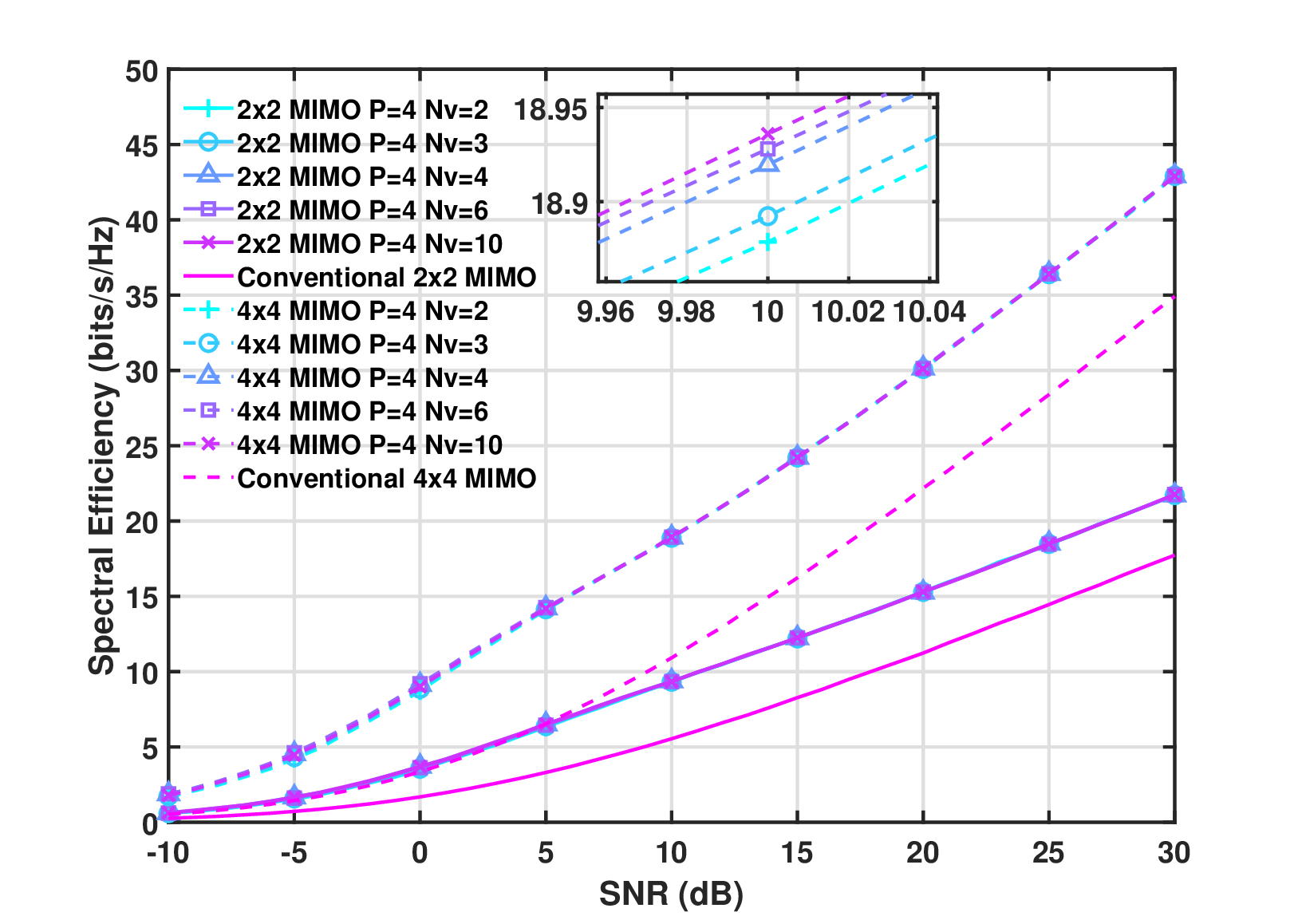}
\par\end{centering}
\begin{centering}
(a)
\par\end{centering}
\begin{centering}
\includegraphics[width=8.5cm]{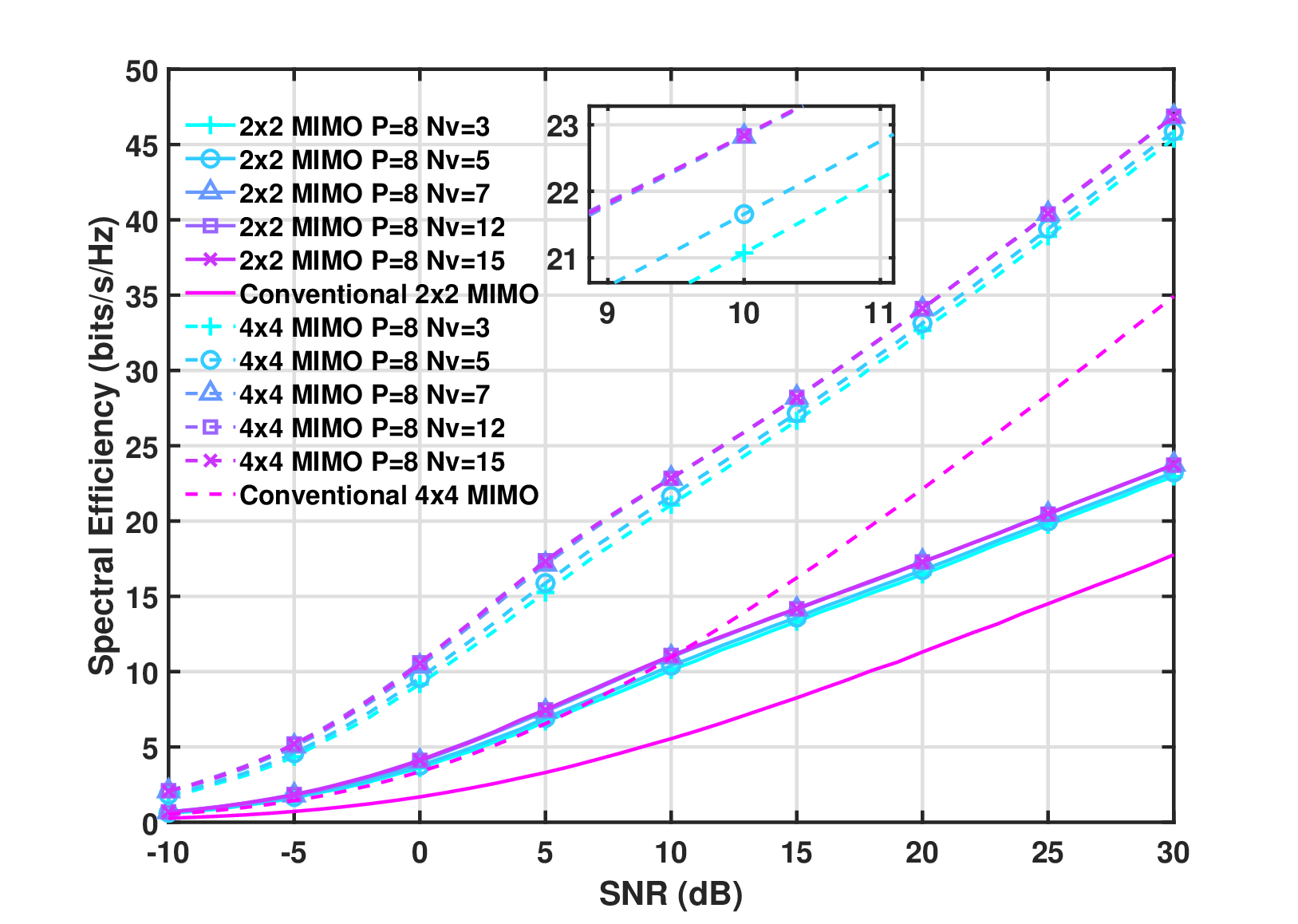}
\par\end{centering}
\begin{centering}
(b)
\par\end{centering}
\caption{Spectral efficiency of the MIMO system with antenna coding when (a)
$P=4$ and (b) $P=8$ antenna coders are used. \label{fig:SE 4}}
\end{figure}
To demonstrate the proposed antenna coding technique for spectral
efficiency enhancement, we utilize the radiation patterns with optimized
codebook, as illustrated in Fig. \ref{fig:Pattern}, to evaluate the
performance of MIMO system with antenna coding. To investigate the
relationship between spectral efficiency and the number of RF switches,
we set different stopping threshold $\Delta g$ in Algorithm 1 to
obtain optimal codebook designs with different number of RF switches
$N_{v}$. The pixel antennas described previously are used at the
transmitter where each pixel antenna is connected to an RF chain.
The conventional spatially isolated antennas are used at the receiver
side. The simulated spectral efficiency, using the derived MI in Section
III.B, of $2\times2$ and $4\times4$ MIMO systems for $P=4$ and
8 cases are shown in Fig. \ref{fig:SE 4}(a) and (b), which are also
benchmarked with results from conventional MIMO system using fixed
antenna configuration. We can make the following observations.

\textit{Firstly, }it can be straightforwardly observed that by using
antenna coding, spectral efficiency of MIMO system can be significantly
enhanced when compared to conventional MIMO system with fixed antenna
configuration. In particular, the SNR gap between the conventional
$4\times4$ MIMO system and the $4\times4$ MIMO system using pixel
antennas with $P=8$ antenna coders and $N_{v}=15$ RF switches is
around 10 dB. In other words, the output power of power amplifier
(PA) can be reduced by 90\% using antenna coding to achieve same spectral
efficiency. In addition, when SNR is 30 dB, the spectral efficiency
of the $4\times4$ MIMO system with antenna coding is 12 bits/s/Hz
higher than the conventional MIMO system, demonstrating the effectiveness
of antenna coding by pixel antenna on enhancing spectral efficiency
of MIMO system.

\textit{Secondly,} we can observe that when using antenna coding,
the spectral efficiency of $4\times4$ MIMO system is twice that of
$2\times2$ MIMO system, which is consistent with the relationship
between the spectral efficiency of conventional $4\times4$ and $2\times2$
MIMO systems. This verifies that antenna coding provides additional
DoF in the beamspace domain for spectral efficiency enhancement without
affecting the original spatial multiplexing of MIMO system.

\textit{Thirdly, }it can be noticed that using more RF switches in
the pixel antenna achieves higher spectral efficiency performance.
This is because the mean correlation among pattern coders decreases
as the number of RF switches increases, indicating that spectral efficiency
can be maximized when the pattern coders are perfectly orthogonal.
However, the increase of spectral efficiency becomes marginal when
the number RF switches is increased to a certain level, which is because
the mean correlation is already small enough for near orthogonality.
In addition, the energy consumption and implementation complexity
can also be higher when more RF switches are used, which will be shown
in the next subsection.

\textit{Fourthly, }comparing spectral efficiency results between $P=4$
and 8 cases in Fig. \ref{fig:SE 4}(a) and (b), we can observe that
the spectral efficiency improvement is marginal by using more RF switches
in $P=4$ case, while spectral efficiency can be effectively enhanced
by using more RF switches in $P=8$ case. This is because the EADoF
of the proposed pixel antenna is 8 so that it is more difficult to
achieve 8 orthogonal pattern coders with limited number of RF switches.
In addition, by combining the results of minimum mean correlation
in Fig. \ref{fig:Obj}, we can find that the spectral efficiency enhancement
is marginal when the mean correlation is lower than 0.1.

\begin{figure}[t]
\begin{centering}
\includegraphics[width=8.5cm]{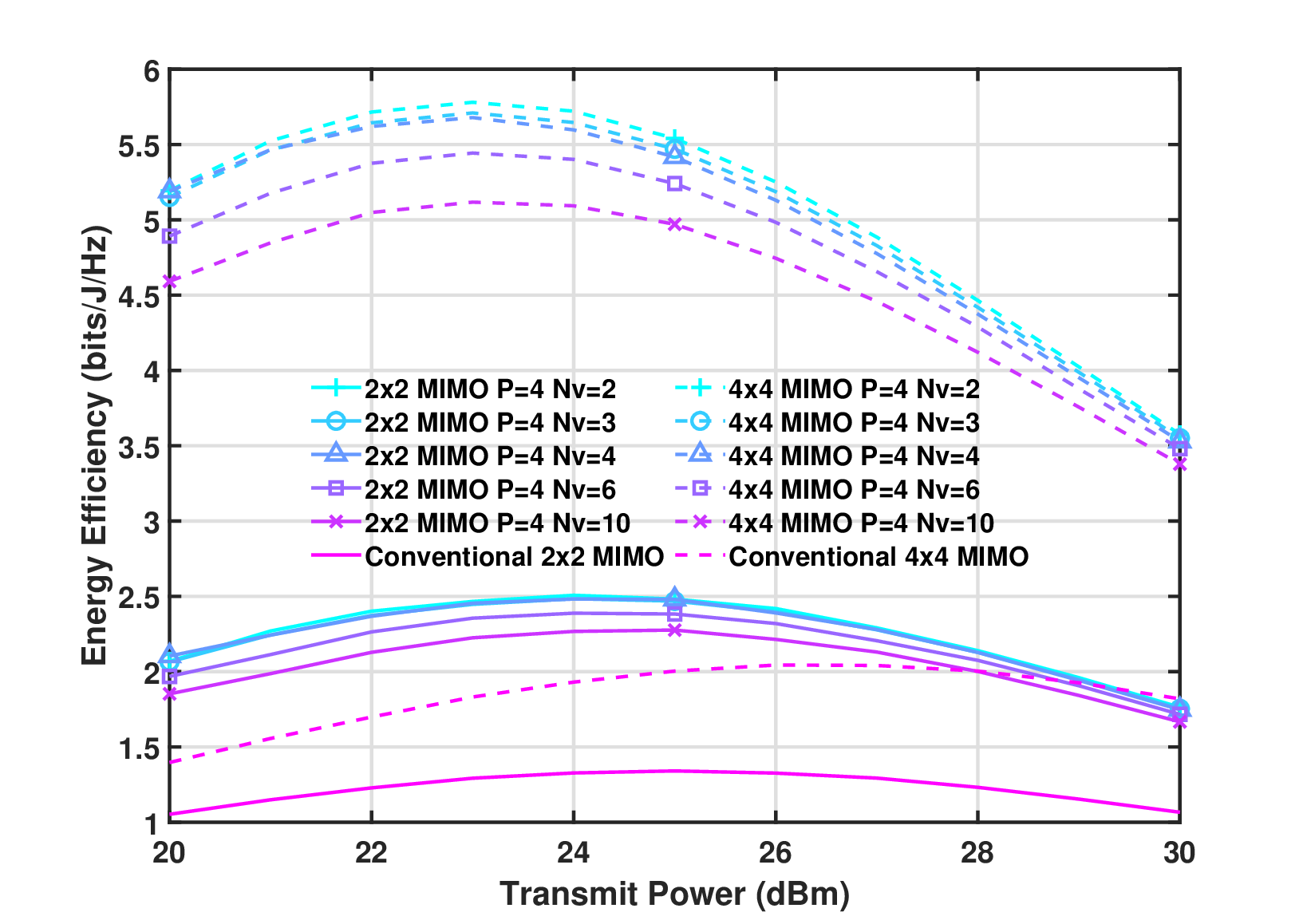}
\par\end{centering}
\begin{centering}
(a)
\par\end{centering}
\begin{centering}
\includegraphics[width=8.5cm]{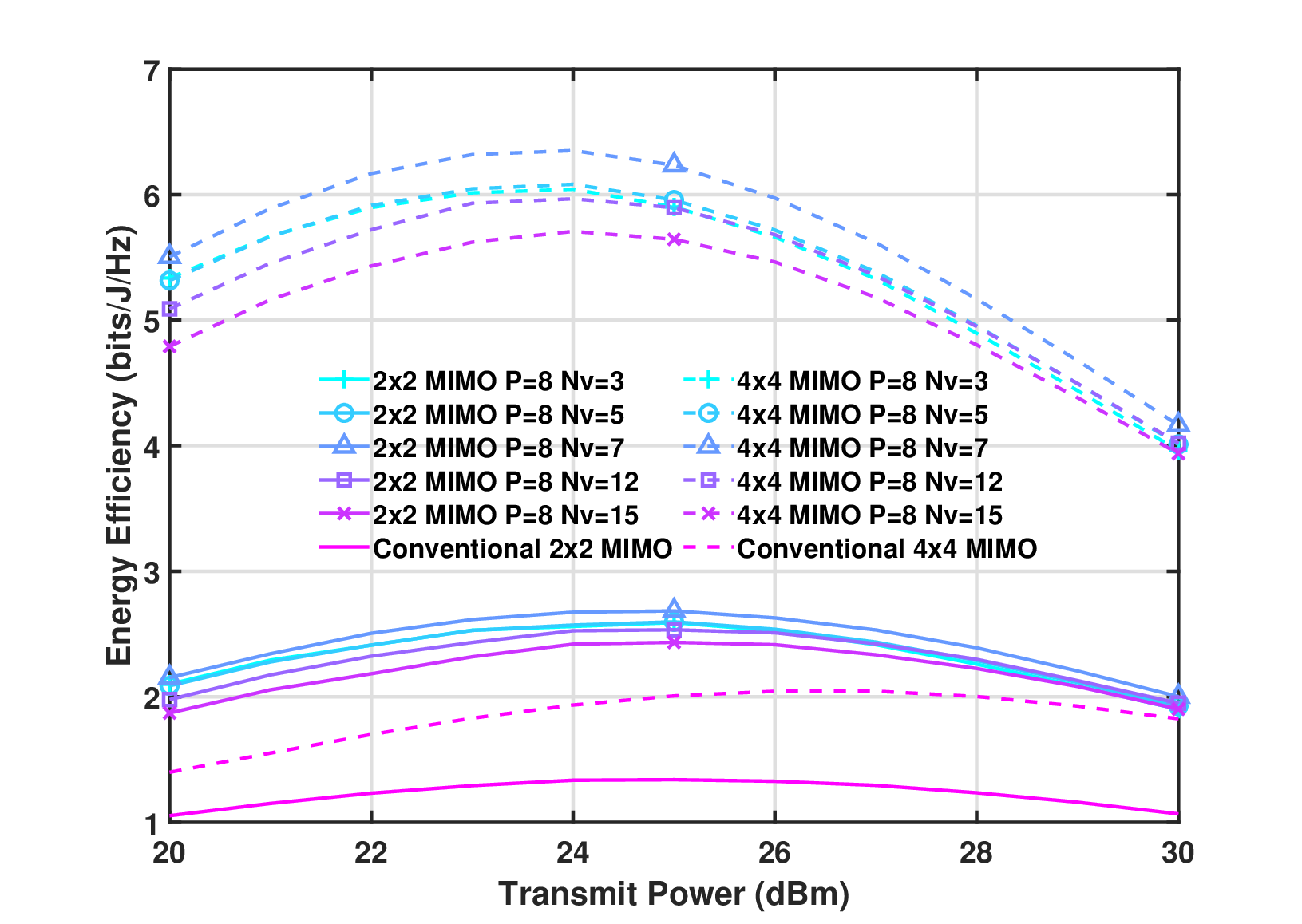}
\par\end{centering}
\begin{centering}
(b)
\par\end{centering}
\caption{Energy efficiency of the MIMO system with antenna coding when (a)
$P=4$ and (b) $P=8$ antenna coders are used. \label{fig:EE 4}}
\end{figure}
To show the benefit of minimizing the number of RF switches in antenna
coding design, it is also important to evaluate the energy efficiency
of MIMO system using pixel antennas. Following the previous approach
in \cite{auer2011much}, we define energy efficiency (EE) as the ratio
of spectral efficiency (SE) to the total power consumption, written
as 
\begin{equation}
\mathrm{EE}=\frac{\mathrm{SE}}{N_{\mathrm{RF}}P_{\mathrm{RF}}+P_{\mathrm{BB}}+P_{\mathrm{PA}}+N_{\mathrm{RF}}N_{\mathrm{SW}}P_{\mathrm{SW}}},\label{eq:EE}
\end{equation}
where $N_{\mathrm{RF}}$ is number of RF chains, $P_{\mathrm{RF}}$
is the RF circuit power consumption for each RF chain, $P_{\mathrm{BB}}$
is the power consumption for baseband processing, $P_{\mathrm{PA}}$
is the power consumption for PA given by $P_{\mathrm{PA}}=\frac{P_{T}}{\eta}$
where $\eta$ is the PA efficiency, $N_{\mathrm{SW}}$ is the number
of RF switches, and $P_{\mathrm{SW}}$ is the power consumption for
each RF switch including switch power dissipation and control circuit
power consumption (both are zero in conventional MIMO system). In
the simulation, the parameters are set as $\eta=20\%$, $P_{\mathrm{BB}}=0.4$
W, $P_{\mathrm{RF}}=0.4$ W, and $P_{\mathrm{SW}}=0.02$ W. The energy
efficiency results are shown in Fig. \ref{fig:EE 4} where we can
make the following observations.

\textit{Firstly, }we can observe again that by using antenna coding,
energy efficiency of MIMO systems are significantly improved when
compared to conventional MIMO system with fixed antenna configuration.
Particularly, energy efficiency for MIMO systems with antenna coding
are around 2 and 3 times higher than that for the conventional MIMO
system in $P=4$ and $P=8$ cases, respectively, demonstrating that
antenna coding can also enhance energy efficiency of MIMO system.

\textit{Secondly,} for the $P=4$ case as shown in Fig. \ref{fig:EE 4}(a),
we can find that energy efficiency increases as the number of RF switches
decreases. This is because the spectral efficiency enhancement by
using more RF switches is marginal while the increase of power consumption
by RF switches is significant, which leads to the degradation of energy
efficiency.

\textit{Thirdly, }for the $P=8$ case as shown in Fig. \ref{fig:EE 4}(b),
it can be observed that using only $\textrm{log}_{2}P=3$ RF switches
does not achieve maximum energy efficiency while using 7 RF switches
achieves the highest energy efficiency performance over the other
RF switch numbers. The reason is that the spectral efficiency performance
gaps between using 3 and 7 RF switches are large. Therefore, a tradeoff
can be performed between the number of RF switches and the energy
efficiency of MIMO system with antenna coding.

\begin{figure}[t]
\begin{centering}
\includegraphics[width=8.5cm]{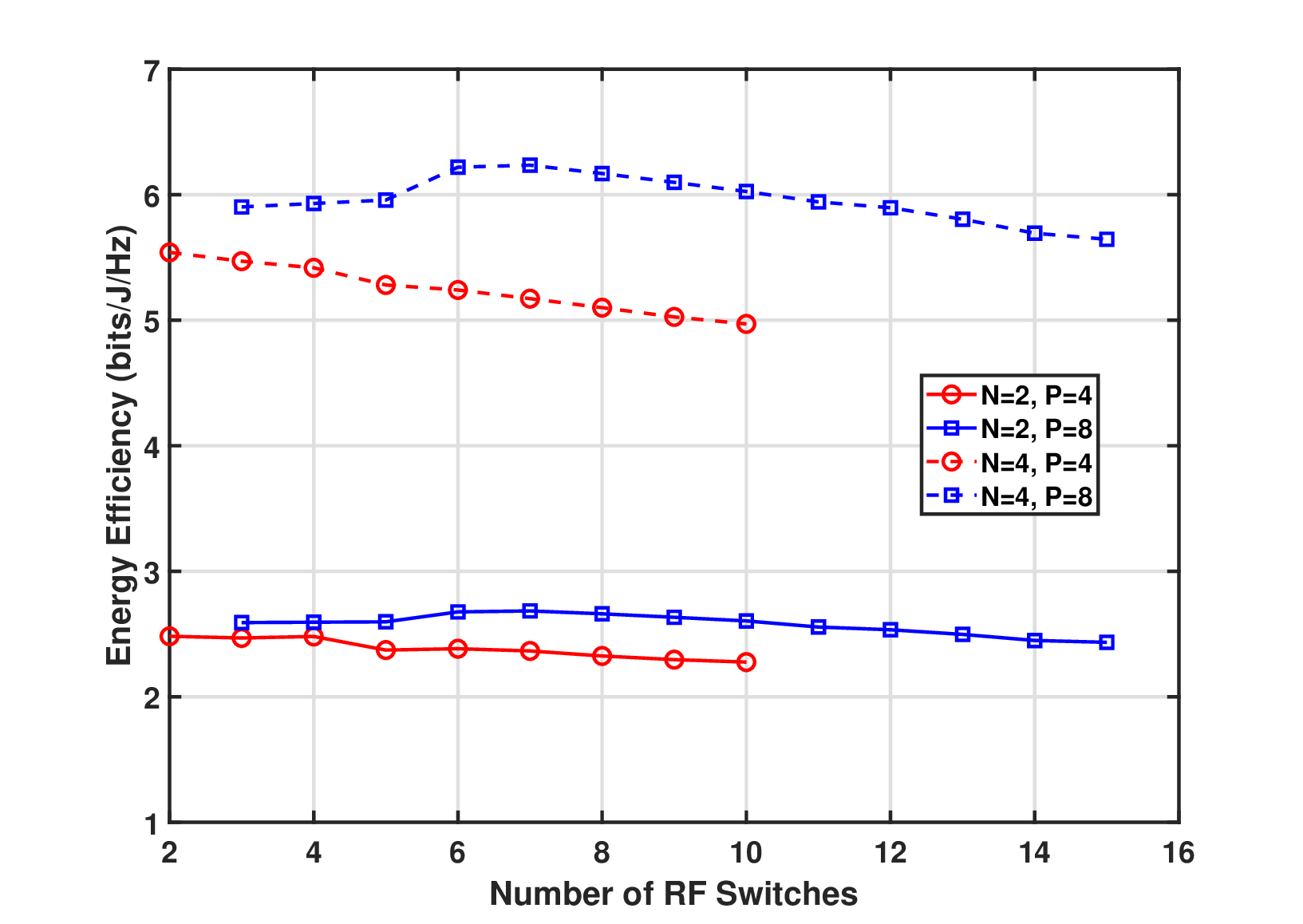}
\par\end{centering}
\caption{Energy efficiency of the MIMO system with antenna coding at $P_{T}=25$
dBm when different number of RF switches are used. \label{fig:EE Sweep}}
\end{figure}
To find the optimal pixel antenna configuration with the maximum energy
efficiency, we also provide the energy efficiency results, as shown
in Fig. \ref{fig:EE Sweep}, when different numbers of RF switches
are used. The transmit power is selected as $P_{T}=25$ dBm. Two observations
can be made.

\textit{Firstly, }we can observe that the maximum energy efficiency
can be achieved when using 7 RF switches for $P=8$ antenna coders
and using 2 RF switches for $P=4$ antenna coders. Therefore, additional
RF switches over $\textrm{log}_{2}P$ are necessary when the spectral
efficiency enhancement by using more RF switches is significant, as
shown in $P=8$ case.

\textit{Secondly,} it can be observed that given the same number of
RF chains, energy efficiencies of MIMO with $P=8$ antenna coders
are higher than that with $P=4$ antenna coders, showing that the
spectral efficiency enhancement is dominant in the energy efficiency
enhancement even though more power are consumed by more RF switches.

To conclude, by using pixel antennas to perform antenna coding, both
spectral efficiency and energy efficiency of MIMO system can be enhanced.
The analysis above can provide a guidance for pixel antenna design
and optimization in antenna coding.

\section{Conclusion}

In this paper, we propose an antenna coding approach for exploiting
the spatial multiplexing capability of pixel antennas. Utilizing the
high reconfigurability of pixel antennas, the radiation patterns of
antenna elements in MIMO systems can be flexibly adjusted so that
additional information bits can be transmitted by modulating the radiation
patterns in the beamspace domain to enhance spectral efficiency.

Specifically in this work, by introducing the antenna coder of the
pixel antenna as the states of RF switches connected between adjacent
pixels, we define the pattern coder of the pixel antenna by decomposing
its radiation pattern into a set of orthonormal basis radiation pattern
in beamspace. Then we introduce the MIMO system model with antenna
coding and derive the spectral efficiency expressions of this system
where the antenna coding for radiation pattern selection contribute
to the spectral efficiency enhancement. Moreover, utilizing the antenna
coder, we analyze the radiation pattern and EADoF of the pixel antenna.
We also formulate the antenna coding optimization to design a codebook
to implement the multiple orthonormal basis radiation patterns for
spectral efficiency enhancement. Further, to reduce the circuit complexity,
we propose an efficient algorithm to minimize the number of RF switches
while maintaining the orthogonality among different pattern coders.

In the numerical simulations, we evaluate the performance of the proposed
optimization algorithm in terms of mean correlation among the optimized
pattern coders, where the mean correlation becomes larger as the number
of RF switches decreases. We also simulate the spectral efficiency
and energy efficiency of the MIMO system with antenna coding using
the pixel antenna. It is shown that the proposed technique using pixel
antennas can enhance spectral efficiency of $4\times4$ MIMO by up
to 12 bits/s/Hz or equivalently reduce the required transmit power
by up to 90\% when compared with conventional MIMO system with fixed
antenna configuration. In addition, a tradeoff can be performed between
the RF switch number and energy efficiency of MIMO systems with antenna
coding. These results demonstrate that the proposed technique can
effectively exploit the spatial multiplexing to enhance the spectral
efficiency of MIMO systems and show the promise of implementing this
technique in upcoming 6G wireless communication.

\bibliographystyle{IEEEtran}

\end{document}